\documentclass[]{interact}

\usepackage[doublespacing]{setspace}
\usepackage{epstopdf}

\usepackage{natbib}
\bibpunct[, ]{(}{)}{;}{a}{}{,}
\renewcommand\bibfont{\fontsize{10}{12}\selectfont}

\theoremstyle{plain}

\theoremstyle{definition}

\theoremstyle{remark}

\newcommand{\lb}{\left (}
\newcommand{\rb}{\right )}

\newcommand{\bx}{\bm x}

\newcommand{\bth}{\bm \theta}

\newcommand{\bphi}{\bm \phi}

\newcommand{\bbet}{\bm \beta}

\usepackage{amsmath}
\usepackage{flexisym}
\usepackage{breqn} 

\usepackage{comment}
\usepackage{algorithm2e}
\usepackage[normalem]{ulem}
\usepackage{mathtools}
\usepackage{tikz}
\usetikzlibrary{shapes}
\usepackage{pgflibraryarrows}
\usepackage{pgflibrarysnakes}
\usepackage{pgfplots}

\usepackage[font=scriptsize, labelfont=bf]{caption} 
\usepackage{subfig}

\begin{document}


\title{Modeling Insurance Claims using Bayesian Nonparametric Regression}

\author{
\name{Mostafa Shams Esfand Abadi\textsuperscript{a} and Kaushik Ghosh\textsuperscript{b}}
\affil{\textsuperscript{a}Department of Statistical Sciences, Wake Forest University, Winston-Salem, NC, USA; \textsuperscript{b}Department of Mathematical Sciences, University of Nevada, Las Vegas, NV, USA}
}

\maketitle

\begin{abstract}
\indent The prediction of future insurance claims based on observed risk factors, or covariates, help the actuary set insurance premiums. Typically, actuaries use parametric regression models to predict claims based on the covariate information. Such models assume the same functional form tying the response to the covariates for each data point. These models are not flexible enough and can fail to accurately capture at the individual level, the relationship between the covariates and the claims frequency and severity, which are often multimodal, highly skewed, and heavy-tailed. In this article, we explore the use of Bayesian nonparametric (BNP) regression models to predict claims frequency and severity based on covariates. In particular, we model claims frequency as a mixture of Poisson regression, and the logarithm of claims severity as a mixture of normal regression. We use the Dirichlet process (DP) and Pitman-Yor process (PY) as a prior for the mixing distribution over the regression parameters. Unlike parametric regression, such models allow each data point to have its individual parameters, making them highly flexible, resulting in improved prediction accuracy. We describe model fitting using MCMC and illustrate their applicability using French motor insurance claims data.

\end{abstract}


\begin{keywords}
Bayesian nonparametric regression; Dirichlet process mixture model; Pitman-Yor process mixture model; Insurance claims modeling. 
\end{keywords}

\section{Introduction}
\indent Insurance claims datasets contain two parts, the claims frequency indicating the number of claims, and the claims severity indicating the monetary amount of each claim. In modeling insurance claims, frequency and severity are often modeled separately. Modeling of the aggregate claim is accomplished by combining these two components. Actuaries often use the standard non-Bayesian finite-dimensional parametric models, such as Poisson regression and normal regression (multiple linear regression), to model the frequency and logarithm of severity respectively. See, for example \cite{tse_2009}, \cite{frees_2009}, and \cite{frees2018loss}. Due to the nature of insurance claims data which are often multimodal, highly skewed to the right, and heavy-tailed, the classical non-Bayesian parametric models are not flexible enough to be able to accurately capture the distribution of claims.

\indent In addition to the lack of flexibility issues, there are some potential biases that can result from considering parametric models, which are model misspecification biases. These biases, which have been discussed in \cite{hong2016}, are serious issues in the case of using finite-dimensional parametric models to predict future insurance claims. Selecting a wrong parametric model may cause serious model misspecification biases, especially for the prediction purpose where the predictive distribution from a misspecified model will have a wrong shape and the predictions will be misleading. To avoid these issues, we need a single flexible model that contains all candidate parametric models, which results in considering the insurance claim distribution itself as an unknown parameter. A Bayesian nonparametric (BNP) model considers the distribution as an unknown parameter and provides an assessment of the actuary's uncertainty for prediction. 

\indent We often want to model insurance claims as a function of covariates, which leads to a regression problem. One issue with parametric regression models is that they assume a fixed response to covariates, and each data point has the same regression parameters. Bayesian nonparametric regression models on the other hand, allow each data point to have its own regression parameters. 

\indent Research on the application of Bayesian nonparametric regression for insurance loss data is limited. \cite{hong2017flexible,hong2018dirichlet} have used a Dirichlet process mixture of log-normals for loss data. Dirichlet process mixture models for insurance loss data has also been discussed in \cite{fellingham2015bayesian}, \cite{richardson2018bayesian}, and \cite{huang2020bayesian}. However, \cite{hong2017flexible,hong2018dirichlet} and \cite{fellingham2015bayesian} have not considered a regression model and the effect of covariates on insurance claims in their studies and have focused on the problem of distribution fitting. \cite{richardson2018bayesian} and \cite{huang2020bayesian} have added the effect of covariates in their research and considered a regression model, however, they have focused only on the claim amounts where the response variable is a continuous variable and the probability base measure of DPMM is a conjugate prior to the likelihood of the model. Complications arise when we are modeling the number of claims, which are count data, and we do not have conjugacy in this case because the base probability measure in our DPMM model for the number of claims is a multivariate normal distribution and is not conjugate to Poisson regression.

\indent In this article, we consider regression models for the effect of covariates on claims frequency and severity, and we use these models to predict the future insurance losses using the techniques from Bayesian nonparametric regression models such as the Dirichlet process mixture model (DPMM) and the Pitman-Yor process mixture model (PYMM), also know as the two-parameter Poisson-Dirichlet process mixture model. We calculate the posterior predictive distribution for claims frequency and severity using the P\'{o}lya urn predictive rule of the Dirichlet process and the Pitman-Yor process. Markov chain Monte Carlo (MCMC) methods, such as Neal (2000)’s Algorithm 8 are used to sample from the posterior distributions in the DPMM and PYMM. One important by-product of these models is the clustering information, which can be used to ascertain the number of mixture components and cluster the observed data into similarity groups. We use the French motor insurance claims data to demonstrate the accuracy and applicability of the proposed models in real data situations. In addition, we discuss the challenges of sampling from the BNP regression models and future direction for this research. The R codes containing the implementation of MCMC algorithms, the calculation of the posterior predictive distribution for DPMM and PYMM, and their clustering performance are available upon request. 

\indent The remainder of this article is organized as follows. In Section 2, we provide a review of two commonly used BNP models, the Dirichlet process mixture model and the Pitman-Yor process mixture model. Then, in Section 3, we present the model for predicting claims frequency. In Section 4, we present the model for claims severity. Finally, in Section 5, we present our conclusions and future directions. 

\indent This article is based on my PhD dissertation \cite[][]{shams2022dissertation}.

\section{Bayesian Nonparametric Models}
\subsection{Dirichlet Process Mixture Models (DPMM)}\label{sec:dpmm}
\indent The Dirichlet process (DP), introduced by \cite{ferguson1973}, provides a nonparametric prior for random distributions. It is denoted by $\text{DP}(\alpha,G_0)$ and has two parameters: a scalar precision parameter $\alpha > 0$ and a base probability measure $G_0$. As a consequence of the stick-breaking representation, proved by \cite{sethuraman1994constructive}, the Dirichlet process generates distributions that are discrete with probability $1$. To model continuous phenomena, this limitation of the DP can be fixed by using the DP as a prior for the mixing distribution over the parameters of a distribution. This leads to Dirichlet process mixture models that can be written as a hierarchical model: 
\begin{align*}
     & y_i \mid \theta_i  \stackrel{ind.}{\sim} F(\theta_i) \nonumber \\
     & \theta_i \mid G  \stackrel{iid}{\sim}  G  \nonumber \\
     & G \sim   \text{DP}(\alpha,G_0). 
\end{align*} 

\indent From a computational point of view, \cite{neal2000markov} has presented several MCMC algorithms for sampling from the posterior distribution of a DPMM model. If we integrate over $G$ in the BNP model above, we obtain the following P\'{o}lya urn predictive rule:
\begin{equation} \label{eq:polya-dpmm}
    \theta_i \mid \theta^{(-i)} \sim \frac{1}{n-1+\alpha} \sum_{j\neq i}^{} \delta_{\theta_j}(.) + \frac{\alpha}{n-1+\alpha} G_0,
\end{equation}
where $\theta^{(-i)} = (\theta_j : j\neq i)$. This leads to a Gibbs sampling algorithm to draw posterior samples in DPMM. When the base probability measure $G_0$ is non-conjugate for the likelihood of the model, the Gibbs sampling is not computationally feasible. \cite{neal2000markov} presents a Gibbs sampling method with auxiliary variables for models with a non-conjugate prior called Neal’s Algorithm 8. This sampling method can also be used for models with a conjugate prior in order to avoid computing the integrals $\int G_0(\theta) f(y_i \mid \theta) \,d\theta$ for the conditional probabilities.

\indent The critical parameter for DPMM is the precision parameter $\alpha$ of the DP prior, controlling the variance and the level of clustering, indicating that the larger $\alpha$ results in $G$ which is closer to the parametric base distribution $G_0$ and therefore larger number of clusters. In this article, we want to apply a full Bayesian model to our data and place a gamma prior on the precision parameter $\alpha$ in the DPMM model. \cite{antoniak1974} has obtained the conditional distribution of the number of distinct components, $k$, given the precision parameter $\alpha$ for the DPMM model. Using this conditional distribution, \cite{escobar1995} has developed a clever way to update $\alpha$ at each iteration. We use this approach to update $\alpha$ for our DPMM model. 

\subsection{Pitman-Yor Process Mixture Models (PYMM)}\label{sec:pym}
\indent The Pitman-Yor process (PY) denoted by $\text{PY}(d,\alpha,G_0)$ is a generalization of the Dirichlet process that has more flexibility over tail behavior than the Dirichlet process which has exponential tails. It has three parameters which are the discount parameter $0 \leq d < 1$, the strength parameter $\alpha > -d$, and the base probability measure $G_0$. Setting $d=0$, the Pitman-Yor process becomes $\text{DP}(\alpha,G_0)$. Pitman-Yor process has a heavier tail than the Dirichlet process. Since many real-world distributions such as insurance loss data have distributions with heavier tails than exponential, this makes the Pitman-Yor process be possibly a better choice for a prior over mixing distribution in mixture models than Dirichlet process \cite[see][]{pitman-yor97,Ishwaran2001,Teh06ahierarchical,fall:hal-00740770}. \par

\indent Similar to Dirichlet process, Pitman-Yor process generates distributions that are discrete with probability $1$. This limitation can be fixed by using the PY as a prior for the mixing distribution over the parameters of one distribution. This leads to the Pitman-Yor process mixture models, also known as the two-parameter Poisson-Dirichlet process mixture models, that can be written as a hierarchical model:
\begin{align*}
     & y_i \mid \theta_i  \stackrel{ind.}{\sim} F(\theta_i) \nonumber \\
     & \theta_i \mid G  \stackrel{iid}{\sim}  G  \nonumber \\
     & G \sim   \text{PY}(d,\alpha,G_0). 
\end{align*}  

Similar to DP, if we integrate over $G$ in the BNP model above, we obtain the following P\'{o}lya urn predictive rule for PY, which can be used to develop a Gibbs sampling algorithm to draw posterior samples in PYMM \cite[see][]{fall:hal-00740770,Ishwaran2001}.
\begin{equation} \label{eq:polya-pym}
    \theta_i \mid \theta^{(-i)} \sim \sum_{j=1}^{K_{n-1}} \frac{n_j-d}{n-1+\alpha} \delta_{\theta_j^{*}}(.) + \frac{\alpha+d K_{n-1}}{n-1+\alpha} G_0,
\end{equation}
where $\theta^{(-i)} = (\theta_j , j\neq i)$, $K_{n-1}$ is the number of distinct components among $\theta^{(-i)}$, and $\theta_j^{*}$ \ $(j=1,\dots,K_{n-1})$ are the unique values among $\theta^{(-i)}$, and $n_j$ is the frequency of $\theta_j^{*}$.

\indent The MCMC algorithm to sample from the PYMM model will be similar to the Neal's Algorithm 8 for the DPMM model. However, there are some differences which are the conditional probabilities in the first step of the Neal's algorithm. According to the P\'{o}lya urn predictive rule for the PYMM model, the configuration cluster label $c_i$ in the first step of the Neal's algorithm is updated by drawing new values from $\{ 1,\dots,h \}$ using the conditional probabilities below \cite[see][]{fall:hal-00740770}.
    \begin{align*}
    P(c_i=c \mid c^{(-i)},y_i, \bm{\phi}_1,\dots,\bm{\phi}_h) \ \propto \
    \begin{cases*}
    (n_{-i,c} - d) F(y_i, \bm{\phi}_c) & \text{for \quad $1 \leq c \leq K_n^{-}$} \\
    ((\alpha + d K_n^{-}) / m) F(y_i, \bm{\phi}_c) & \text{for \quad $K_n^{-} < c \leq h$,} 
    \end{cases*}
    \end{align*}
where $n_{-i,c}$ is the number of $c_j$ for $j \neq i$ that are equal to $c$ and $K_n^{-}$ is the number of current distinct components among the observations except the observation $i$.

\indent The discount parameter $d$ and strength parameter $\alpha$ in the PYMM model are critical parameters for this model. In this article, we place a uniform prior on the discount parameter $d$ and a log-normal prior on the sum of the strength parameter and discount parameter, $(\alpha+d)$, because it should be nonnegative. Let $K_n$ be a random variable indicating the current number of distinct components in each iteration of the MCMC sampling algorithm. In order to determine the conditional posterior distributions for $d$ and $\alpha$ and then update $d$ and $\alpha$ at each iteration, we use the conditional distribution of $K_n$ given both $d$ and $\alpha$, which is calculated in \cite{lijoi2007}. In order to sample from the conditional posterior distributions for $d$ and $\alpha$, we use the random-walk Metropolis–Hastings algorithm. As stated in \cite{roberts2009}, tuning of associated proposal variances in  random-walk Metropolis–Hastings is crucial to achieve efficient mixing. Therefore, we use the Adaptive Metropolis-within-Gibbs method in \cite{roberts2009} to update proposal variances at each iteration. 

\subsection{Clustering Property of BNP Models} \label{sec:clustering}
As stated in the previous sections, the Dirichlet process and Pitman-Yor process generate distributions that are discrete with probability $1$. Therefore, the probability measure $G$ in the BNP models is a discrete probability measure, and this implies a positive probability for the ties among the parameters $\theta_1,\theta_2,\dots,\theta_n$. We can use these ties to define clusters, i.e. $k < n$ distinct values of $\theta_i$'s, that induces a clustering structure on the dataset. \par

\indent We check the posterior clustering performance of our Bayesian nonparametric regression models in our real data application. Let us assume that there are $n$ data points in our training dataset and our model regression parameter vectors are $\bth_1,\dots,\bth_n$ where $\bth_i$ is the regression parameter vector related to the data point $i$. We do the clustering based on samples from the posterior distribution of $\bth_1,\dots,\bth_n$. Clustering requires some methods for computing the dissimilarity between each pair of observations. We compute the $n \times n$ dissimilarity matrix based on samples from posterior distribution of $\bth_1,\dots,\bth_n$. The element in the $i$th row and $j$th column of this dissimilarity matrix is the proportion of samples from posterior distribution where observations $i$ and $j$ have different coefficient vector, i.e. posterior samples for $\bth_i$ and $\bth_j$ are different. \par

\indent Unlike the traditional clustering algorithms, the DPMMs and PYMMs do not require us to set the number of clusters $k$ in advance. These two models define a mixture model with countably infinitely many components and can infer $k$ from the data and allows $k$ to grow as more data are collected.

\section{Modeling Claims Frequency}
\indent The first step in developing any insurance pricing model is predicting the claims frequency, i.e. the number of claims. An actuary wants to model the claims frequency as a function of some covariates which are called risk factors in insurance literature. Let us assume that there are $n$ independent policyholders with  a set of $k$ risk factors for each. The $i$th policyholder's risk factors are denoted by the vector $\bx_i = (1,x_{i1},\dots,x_{ik})^T$. The $i$th policyholder's recorded number of claims is denoted by $y_i \in \{0,1,2,\dots\}$. In order to determine the size of potential losses in any type of insurance, one must also know the corresponding exposure \citep{frees2018loss}. We let $t_i$ represent the length of time or exposure for the $i$th policyholder. The classical non-Bayesian parametric Poisson regression model that has been widely used to model the number of claims of the $i$th individual is
\begin{align*} 
     y_i \mid \bbet \ \stackrel{ind.}{\sim} \ \text{Poisson}(t_i \exp(\bx'_i \bbet)), 
\end{align*}
where $\bbet = (\beta_{0},\beta_{1},\dots,\beta_{k})^T$ is the vector of regression coefficients. However, the above model assumes that each data point has the same regression parameter vector, $\bbet$, and thus similar response to covariates for each individual. In this section, we use a DPMM and a PYMM model which allow each data point to have its own regression parameter vector. Due to the clustering property of DP prior and PY prior, the data points are then clustered by their shared regression parameters \cite[see][]{hartman2010,hannah11a}.

\subsection{Model Description} \label{sec:bnp-freq}
\indent The BNP Poisson regression parameters here are the regression coefficient vectors, $\bbet_i = (\beta_{i0},\beta_{i1},\dots,\beta_{ik})$, for $i = 1,\dots,n$. We propose two BNP models for the number of claims data as follows. 

\indent First, we set a DP prior on the mixing distribution over the regression parameters, $\bbet_i$, and we model the number of claims, $y_i$, using a Dirichlet process mixture of Poisson regression. Since we want to apply a full Bayesian model to our data, we place a gamma prior on the precision parameter $\alpha$. Our DPMM model can be written as a hierarchical model:
\begin{align} \label{eq:dpmm-freq}
     & y_i \mid \bbet_i \ \stackrel{ind.}{\sim} \ \text{Poisson}(t_i \exp(\bx'_i \bbet_i)), \quad i = 1,\dots,n \nonumber \\
     & \bbet_i \mid G \ \stackrel{iid}{\sim} \ G, \quad i = 1,\dots,n \nonumber \\
     & G \ \sim \ \text{DP}(\alpha,G_0) \\
     & G_0 \ = \ \text{MN}(\bm{0}_{k+1} , I_{k+1}) \nonumber \\
     & \alpha \ \sim \ \text{Gamma}(1,1). \nonumber
\end{align}
Here, the base probability measure $G_0$ is taken to be a multivariate normal distribution with a $(k+1)$-dimensional mean vector $\bm{0}_{k+1}$ and a $(k+1) \times (k+1)$ covariance matrix $I_{k+1}$ where $I_{k+1}$ is the identity matrix of size $k+1$.

\indent Alternatively, we set a PY prior instead of a DP prior on the mixing distribution over the regression parameters, $\bbet_i$, and we model the number of claims, $y_i$, using a Pitman-Yor process mixture of Poisson regression. In this case, we place a uniform prior on the discount parameter $d$ and a log-normal prior on the sum of precision parameter and discount parameter, $(\alpha+d)$, because it should be nonnegative. Our PYMM model can be written as a hierarchical model:
\begin{align} \label{eq:pym-freq}
     & y_i \mid \bbet_i  \stackrel{ind.}{\sim} \ \text{Poisson}(t_i \exp(\bx'_i \bbet_i)), \quad i = 1,\dots,n \nonumber \\
     & \bbet_i \mid G \ \stackrel{iid}{\sim} \ G, \quad i = 1,\dots,n  \nonumber \\
     & G \ \sim  \ \text{PY}(d,\alpha,G_0) \\
     & G_0 \ = \ \text{MN}(\bm{0}_{k+1} , I_{k+1}) \nonumber \\
     & d \sim \text{Uniform}(0,1) \nonumber \\
     & (\alpha+d) \mid d \sim \text{Log-Normal}(0,1). \nonumber
\end{align} 

\subsection{Computation of Posterior Predictive Distributions} \label{sec:pred_freq_dpmm_pym}
\indent In this section, we are interested in the predictive distribution for the future number of claims $y_{n+1}$. This is the conditional probability mass function of $y_{n+1}$, given the data $y_1,\dots,y_n$. In this prediction, we are implicitly assuming that the new covariate vector is $\bm{x}_{n+1}$ and exposure is $t_{n+1}$. This means that whenever we use $f(y_{n+1} \mid y_1,\dots,y_n)$, it is implicitly equivalent to \\
\begin{dmath*}[style={\small}]
    f{(y_{n+1} \mid y_1,\dots,y_n)} = f{(y_{n+1} \mid y_1, \dots, y_n, \bm{x}_{n+1}, t_{n+1})}.\\
\end{dmath*}
Note that the parameters in DPMM model are the coefficients of the BNP Poisson regression, which are $\bbet_i$. It can be shown that \\
\begin{dmath}[style={\small},spread={5pt}] \label{eq:pred-dpmm}
    f{(y_{n+1} \mid y_1,\dots,y_n)} = \lb \int \frac{\alpha}{\alpha+n} f{(\bbet_1,\dots,\bbet_n,\alpha \mid y_1,\dots,y_n)} \,d\bbet_{n} \dots \,d\bbet_{1} \,d\alpha\rb 
    \times \lb \int f{(y_{n+1} \mid \bbet_{n+1})G_0(d\bbet_{n+1})} \rb 
    + \idotsint \frac{1}{\alpha+n} \sum_{j=1}^{n} f{(y_{n+1} \mid \bbet_{j})} 
    \times f{(\bbet_1,\dots,\bbet_n,\alpha \mid y_1,\dots,y_n)} \,d\bbet_{n} \dots \,d\bbet_{1} \,d\alpha. 
\end{dmath}
By getting $T$ samples from the posterior distribution of $\bbet=(\bbet_1,\dots,\bbet_n)$ and $\alpha$, with the $t^{\text{th}}$ sample being $\bbet^{(t)}=(\bbet_1^{(t)},\dots,\bbet_n^{(t)})$ and $\alpha^{(t)}$, for $t=1,\dots,T$, the above can be approximated using \\
\begin{dmath}[style={\small},spread={5pt}] \label{eq:freq1}
    f{(y_{n+1} \mid y_1,\dots,y_n)} \approx \frac{1}{T} \sum_{t=1}^{T} \lb \frac{\alpha^{(t)}}{\alpha^{(t)}+n} \rb \int f{(y_{n+1} \mid \bbet_{n+1})}G_0(d\bbet_{n+1}) 
    + \frac{1}{T} \sum_{t=1}^{T} \lb\frac{1}{\alpha^{(t)}+n} \sum_{j=1}^{n} f{(y_{n+1} \mid \bbet_{j}^{(t)})}\rb. \\
\end{dmath}
Next, we begin to calculate the predictive distribution for our Bayesian nonparametric regression model in (\ref{eq:dpmm-freq}), where in equation (\ref{eq:freq1}) above, $f$ is the pmf of the Poisson distribution and $G_0$ is the multivariate normal distribution with a $(k+1)$-dimensional mean vector $\bbet_0 = \bm{0}_{k+1}$ and a $(k+1) \times (k+1)$ covariance matrix $\Sigma_0 = I_{k+1}$ which is the identity matrix of size $k+1$. This means that \\
\begin{dmath*}[style={\small}]
    f{(y_{n+1} \mid \bbet_{n+1})} = \frac{e^{-t_{n+1}.e^{\left(\bx'_{n+1}\bbet_{n+1}\right)}}.\left(t_{n+1}.e^{\left(\bx'_{n+1}\bbet_{n+1}\right)}\right)^{y_{n+1}}}{y_{n+1}!},\\
\end{dmath*}
where $t_{n+1}$ is the exposure and $\bx_{n+1}$ is a vector of covariates which represents a risk class. For $G_0$, we have \\
\begin{dmath*}[style={\small}]
    g_0(\bbet_{n+1}) = (2\pi)^{\frac{-(k+1)}{2}} e^{\frac{-1}{2}\bbet'_{n+1}\bbet_{n+1}}. \\
\end{dmath*}
We now split the equation (\ref{eq:freq1}) into two parts and calculate each part separately. First, we calculate the following integral: \\
\begin{dmath*}[style={\small},spread={5pt}]
    \int f{(y_{n+1} \mid \bbet_{n+1})} G_0(d\bbet_{n+1}) 
    = \int \frac{e^{-t_{n+1}.e^{\left(\bx'_{n+1}\bbet_{n+1}\right)}}
    \times \left(t_{n+1}.e^{\left(\bx'_{n+1}\bbet_{n+1}\right)}\right)^{y_{n+1}}}
    {y_{n+1}!} \times (2\pi)^{\frac{-(k+1)}{2}} e^{\frac{-1}{2}\bbet'_{n+1}\bbet_{n+1}} \,d\bbet_{n+1}. 
\end{dmath*}
Let $h(\bbet_{n+1})$ denote \\
\begin{dmath*}[style={\small}]
    h(\bbet_{n+1}) = -t_{n+1}.e^{\left(\bx'_{n+1}\bbet_{n+1}\right)}+y_{n+1}\left(\log(t_{n+1})+\bx'_{n+1}\bbet_{n+1}\right)-\frac{1}{2}\bbet'_{n+1}\bbet_{n+1}.\\
\end{dmath*}
Using multivariate Laplace approximation, we have \\
\begin{dmath}[style={\small},spread={5pt}] \label{eq:freq2}
    \int f{(y_{n+1} \mid \bbet_{n+1})}G_0(d\bbet_{n+1}) 
    \approx \frac{(2\pi)^{\frac{-(k+1)}{2}}}{y_{n+1}!} \exp{\left(h(\hat{\bbet}_{n+1})\right)}(2\pi)^{\frac{k+1}{2}} \lvert\hat{\Sigma}\rvert^{\frac{1}{2}}
    \approx \frac{1}{y_{n+1}!} \exp{\left(h(\hat{\bbet}_{n+1})\right)} \lvert\hat{\Sigma}\rvert^{\frac{1}{2}},\\
\end{dmath}
where $\hat{\bbet}_{n+1}$ is the value of $\bbet_{n+1}$ such that $h'(\hat{\bbet}_{n+1})=0$ and $\hat{\Sigma}$ the inverse of the Hessian of $h$ evaluated at $\hat{\bbet}_{n+1}$. Hence, \\
\begin{dmath}[style={\small},spread={5pt}] \label{eq:dpmm-pred-num}
    f{(y_{n+1} \mid y_1,\dots,y_n)} \approx \left(\frac{1}{T} \sum_{t=1}^{T} \frac{\alpha^{(t)}}{\alpha^{(t)}+n}\right) \frac{1}{y_{n+1}!} \exp{\left(h(\hat{\bbet}_{n+1})\right)} {\lvert\hat{\Sigma}\rvert}^{\frac{1}{2}}
    +\frac{1}{T} \sum_{t=1}^{T} \lb \frac{1}{\alpha^{(t)}+n} \sum_{j=1}^{n} \frac{1}{y_{n+1}!} \exp{\left(-t_{n+1}.e^{\left(\bx'_{n+1}\bbet_{j}^{(t)}\right)}+y_{n+1}\left(\log(t_{n+1})+\bx'_{n+1}\bbet_{j}^{(t)}\right)\right)} \rb. 
\end{dmath}

\indent Similarly, the posterior predictive distribution of our PYMM model for claims frequency can be approximated using: \\
\begin{dmath}[style={\small},spread={5pt}]\label{eq:pym-pred-num}
    f{(y_{n+1} \mid y_1,\dots,y_n)} \approx \lb\frac{1}{T} \sum_{t=1}^{T} \frac{\alpha^{(t)}+d^{(t)} K_n^{(t)}}{\alpha^{(t)} + n}\rb \frac{1}{y_{n+1}!} \exp{\left(h(\hat{\bbet}_{n+1})\right)} \lvert\hat{\Sigma}\rvert^{\frac{1}{2}}
    + \frac{1}{T} \sum_{t=1}^{T} \frac{1}{(\alpha^{(t)}+n)} \left( \sum_{j=1}^{n} \frac{1}{y_{n+1}!} \exp{\left(-t_{n+1}.e^{\left(\bx'_{n+1}\bbet_{j}^{(t)}\right)}+y_{n+1}\left(\log(t_{n+1})+\bx'_{n+1}\bbet_{j}^{(t)}\right)\right)} \right.
    \left. \qquad \qquad - d^{(t)} \sum_{j=1}^{K_{n}^{(t)}} \frac{1}{y_{n+1}!} \exp{\left(-t_{n+1}.e^{\left(\bx'_{n+1}{\bbet_j^*}^{(t)}\right)}+y_{n+1}\left(\log(t_{n+1})+\bx'_{n+1}{\bbet_j^*}^{(t)}\right)\right)}\right), \\
\end{dmath}
where $\hat{\bbet}_{n+1}$ is the value of $\bbet_{n+1}$ such that $h'(\hat{\bbet}_{n+1})=0$ and $\hat{\Sigma}$ the inverse of the Hessian of $h$ evaluated at $\hat{\bbet}_{n+1}$ calculated by multivariate Laplace approximation. 

\subsection{Posterior Sampling} \label{sec:implem-freq}
We use Neal’s Algorithm 8 to sample from the posterior distribution of the Bayesian nonparametric regression parameters $\bbet_1,\dots,\bbet_n$ presented in models (\ref{eq:dpmm-freq}) and (\ref{eq:pym-freq}). In the second step of this algorithm, we draw a new value for the distinct parameter $\bm{\phi}_c$ from the posterior distribution $(\bm{\phi} \mid \bm{y_c})$ when we assume that the base measure $G_0$ is the prior distribution. Since the base probability measure $G_0$ in our models (\ref{eq:dpmm-freq}) and (\ref{eq:pym-freq}) is a multivariate normal distribution and is not conjugate to Poisson likelihood, we replace the Gibbs sampling update for $\bphi_c$ by a Metropolis–Hastings update to draw new values for $\bphi_c$ at each iteration. The mean and variance of the multivariate normal proposal distribution are obtained using a Laplace approximation. As explained in the Section \ref{sec:dpmm}, for the DPMM model, we use the approach mentioned in \cite{escobar1995} to update $\alpha$ at each iteration. For the PYMM model, we use our approach explained in the Section \ref{sec:pym} to update $d$ and $\alpha$ at each iteration.

\section{Illustration: French Motor Insurance Claims Frequency Dataset}
\indent In this section, we demonstrate the application of our BNP regression models with a real insurance loss data, the French motor insurance claims dataset which is available as part of the R package ``CASdatasets''. This dataset contains two datasets freMTPLfreq and freMTPLsev where the risk features are collected for $413,169$ motor third-part liability policies (observed mostly during a one year period). In addition, we have claim numbers by policy as well as the corresponding claim amounts. The freMTPLfreq dataset contains the risk features and the claim number, while the freMTPLsev dataset contains the claim amount and the corresponding policy ID \cite[see][]{casdatasets}. In this section, we consider freMTPLfreq dataset which contains claim numbers. The dataset contains $10$ columns as follows:
\begin{itemize}
    \item \textit{PolicyID} The policy ID (used to link with the claims dataset).
    \item \textit{ClaimNb} Number of claims during the exposure period.
    \item \textit{Exposure} The period of exposure for a policy, in years.
    \item \textit{Power} The power of the car (ordered categorical variable with $12$ levels).
    \item \textit{CarAge} The vehicle age, in years.
    \item \textit{DriverAge} The driver age, in years (in France, people can drive a car at 18).
    \item \textit{Brand} The car brand categorized into the following $7$ groups: A- Renault Nissan and Citroen; B- Volkswagen, Audi, Skoda and Seat; C- Opel, General Motors and Ford; D- Fiat; E- Mercedes Chrysler and BMW; F- Japanese (except Nissan) and Korean; G- other.
    \item \textit{Gas} The car fuel type with $2$ levels: Diesel or gasoline.
    \item \textit{Region} The policy region in France with $10$ levels (based on the 1970-2015 classification).
    \item \textit{Density} The density of inhabitants (number of inhabitants per km$^2$) in the city the driver of the car lives in.
\end{itemize}

\indent We considered the number of claims in this dataset as response variable. Due to computational limit and time, we used two covariates in our BNP regression models: driver age (DriverAge column) and car age (CarAge column) which are ordinal variables treated as continuous variables. We randomly selected $2000$ data points as training data points to fit our Bayesian nonparametric regression models, and used the remainder as testing data to evaluate their prediction performance. We ran these models with two covariates for the $2000$ training data points in R for $50000$ MCMC iterations on a HPC cluster, and it took $3.19$ days for the DPMM and $2.91$ days for the PYMM to be completed. Since these categorical independent variables are ordinal, we entered them into our model as continuous independent variables, and not as dummy variables. To avoid overflow or underflow issues in R, we standardized all the covariates. Since we have two covariates here, the risk factor vector is in the form of $\bx_i = (1,x_{i1},x_{i2})^T$, for $i = 1,\dots,n$, and the BNP regression model parameters are the regression coefficients vectors, $\bbet_i = (\beta_{i0},\beta_{i1},\beta_{i2})$, for $i = 1,\dots,n$. Therefore, our Dirichlet process mixture of Poisson regression model here is:
\begin{align*}
     & y_i \mid \bbet_i \ \stackrel{ind.}{\sim} \ \text{Poisson}(t_i \exp(\beta_{i0}+\beta_{i1} x_{i1}+\beta_{i2} x_{i2})) \nonumber \\
     & \bbet_i \mid G \ \stackrel{iid}{\sim} \ G  \nonumber \\
     & G \ \sim \ \text{DP}(\alpha,G_0) \nonumber \\
     & G_0 \ = \ \text{Multivariate-Normal}(\bm{0}_3 , I_3) \nonumber \\
     & \alpha \ \sim \ \text{Gamma}(1,1).
\end{align*}
Similarly, our Pitman-Yor process mixture of Poisson regression model is:
\begin{align*} 
     & y_i \mid \bbet_i \ \stackrel{ind.}{\sim} \ \text{Poisson}(t_i \exp(\beta_{i0}+\beta_{i1} x_{i1}+\beta_{i2} x_{i2})) \nonumber \\
     & \bbet_i \mid G \ \stackrel{iid}{\sim} \ G  \nonumber \\
     & G \ \sim  \ \text{PY}(d,\alpha,G_0) \\
     & G_0 \ = \ \text{Multivariate-Normal}(\bm{0}_3 , I_3) \\
     & d \ \sim \ \text{Uniform}(0,1) \\
     & (\alpha + d) \mid d \ \sim \ \text{Log-Normal}(0,1).
\end{align*}

We ran the algorithm for $50000$ iterations, with a burn-in of first $25000$ iterations. We used trace plots and autocorrelation function plots to assess the mixing of the Markov chain and check the convergence of the Markov chain to stationary distribution. All the plots indicated a good convergence. In addition, diagnostic tests of convergence to the equilibrium distribution of the Markov chain provided by the R package ``coda'' \cite[see][]{coda} all indicated a good convergence. 

\begin{figure}[ht!]
    \centering
    \includegraphics[width=.45\textwidth]{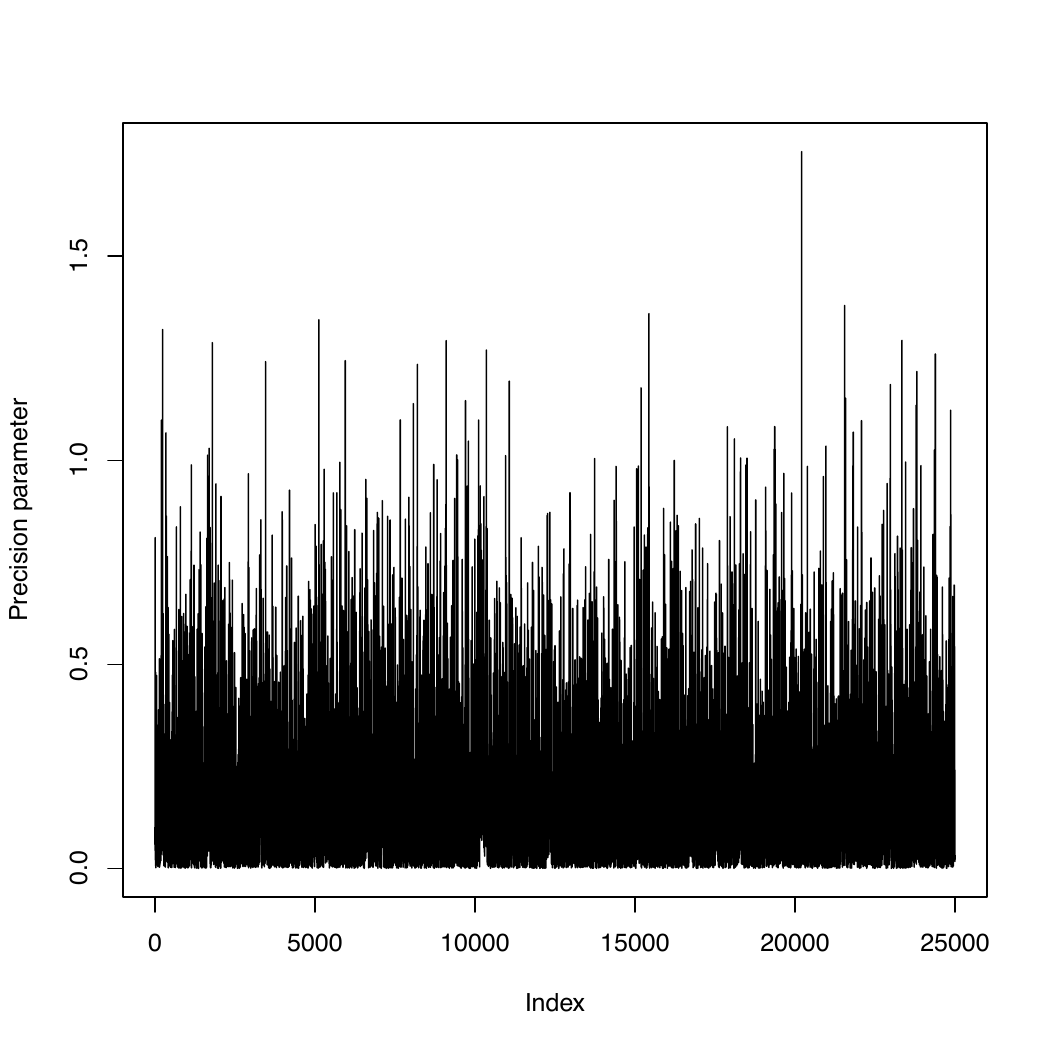}
    \caption{Trace plot of $\alpha$ in DPMM model for the French motor claims frequency dataset.}
    \label{fig:alpha-dpmm-num}
\end{figure}

\begin{figure}[ht!]
    \centering
    \subfloat[Trace plot of $\alpha$]{\includegraphics[width=.45\textwidth]{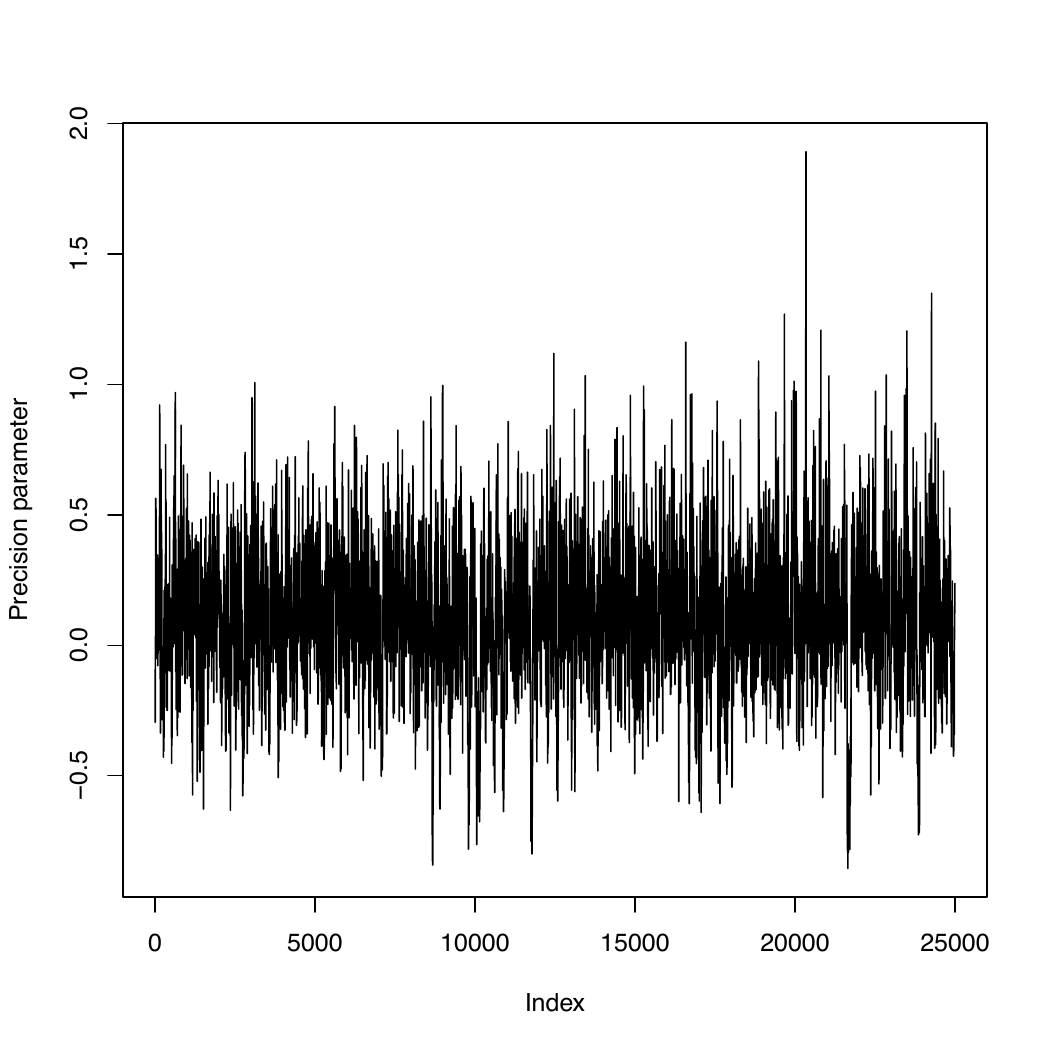}}\quad
    \subfloat[Trace plot of $d$]{\includegraphics[width=.45\textwidth]{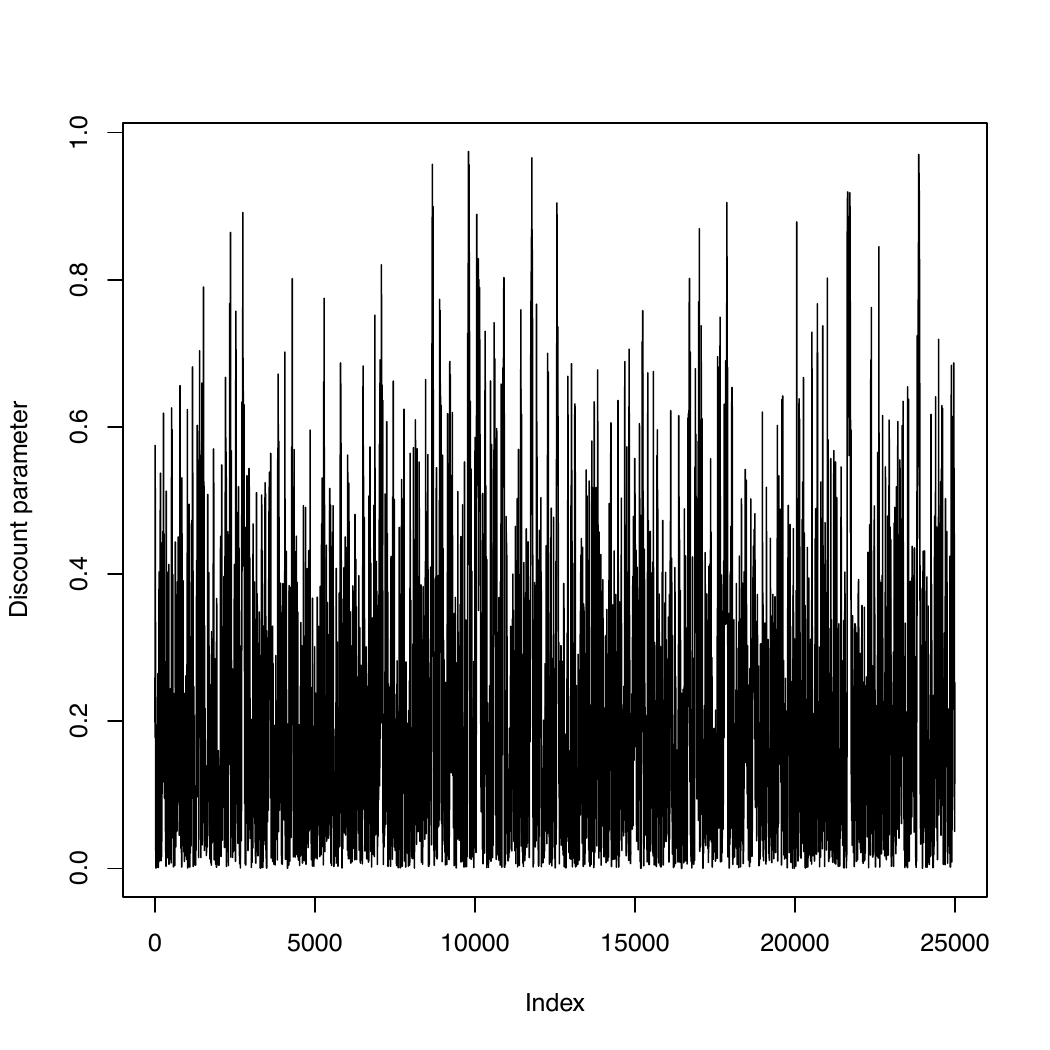}}
    \caption{Trace plots of $\alpha$ and $d$ in PYMM model for the French motor claims frequency dataset.}
    \label{fig:alphad-pym-num}
\end{figure}

\begin{figure}[ht!]
    \centering
    \subfloat[DPMM]{\includegraphics[width=.45\textwidth]{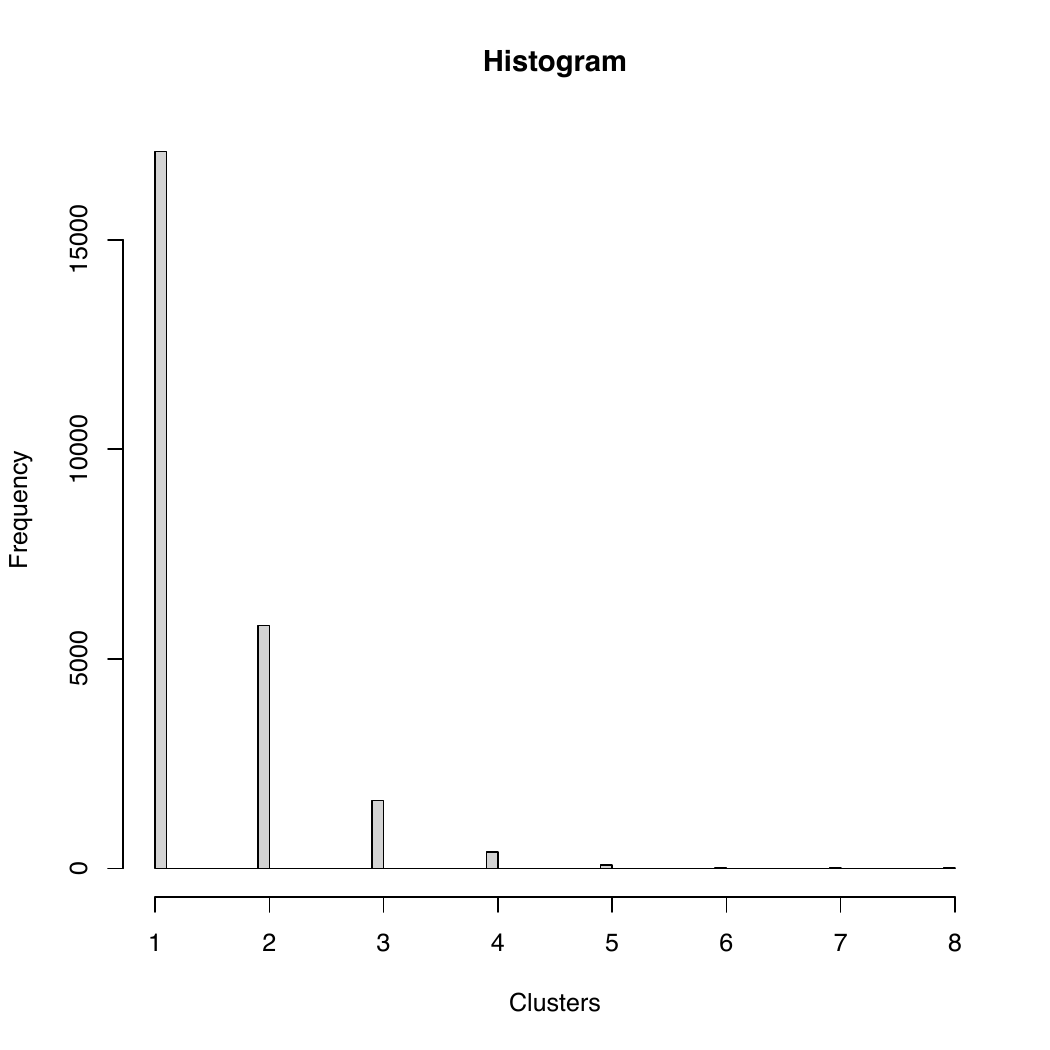}}\quad
    \subfloat[PYMM]{\includegraphics[width=.45\textwidth]{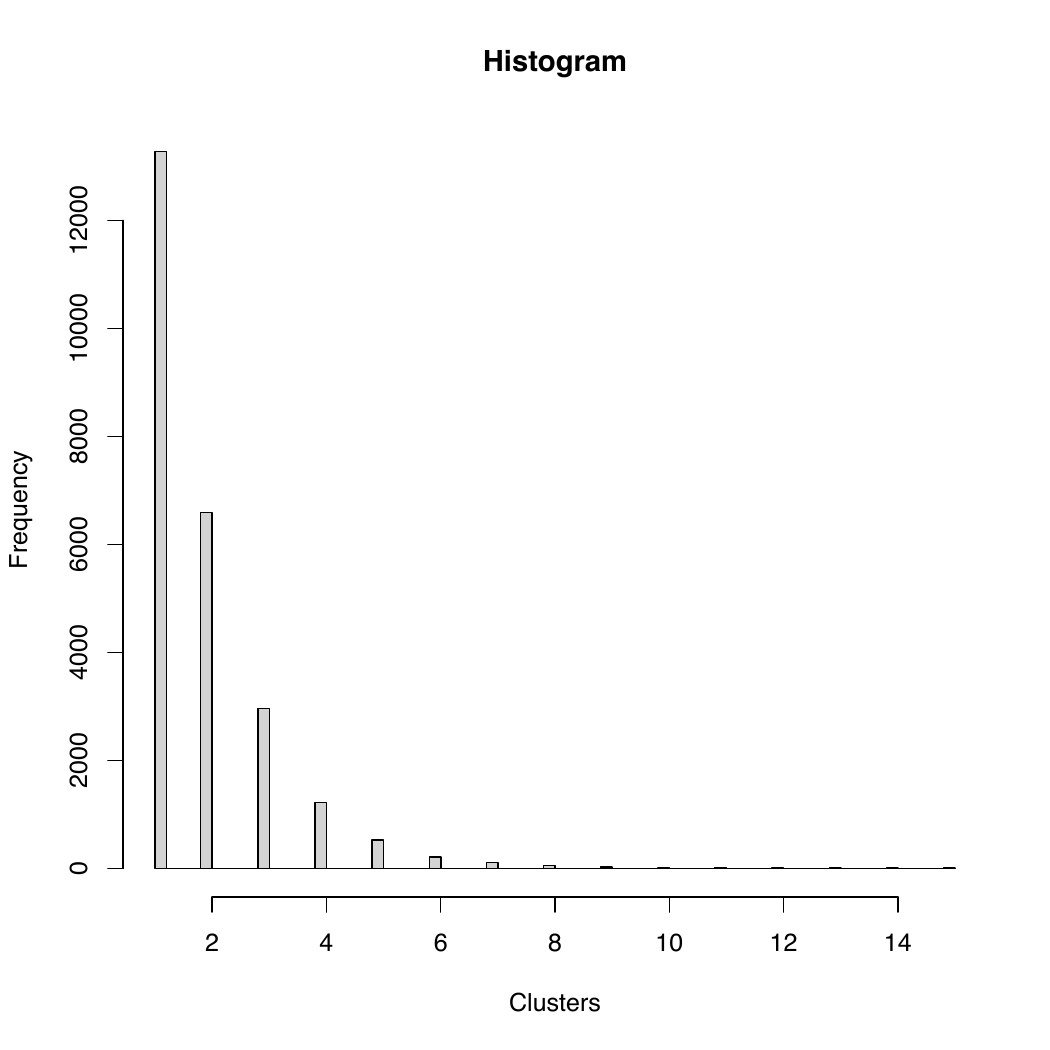}}
    \caption{Histogram of the number of distinct components for the French motor claims frequency dataset.}
    \label{fig:hist-num}
\end{figure}

The posterior sample trace plots for $\alpha$ in the DPMM model and for $\alpha$ and $d$ in the PYMM model are shown in Figures \ref{fig:alpha-dpmm-num} and \ref{fig:alphad-pym-num} respectively. Histograms of the number of distinct components for DPMM and PYMM are shown in Figure \ref{fig:hist-num}. 

\indent Since our main goal in insurance is prediction, we are interested in the predictive distribution for the future number of claims $y_{n+1}$. For illustrative purposes, we calculated the predictive distribution of the number of claims for car age $2$ and driver age $62$ (observed during a one year period) using our Bayesian nonparametric regression models, as described in Equations (\ref{eq:dpmm-pred-num}) and (\ref{eq:pym-pred-num}). We used a chi-square goodness-of-fit test to compare the predictive distribution of the number of claims obtained from the testing data with the observed distribution in the testing data. Since there were $3$ categories of claims for the class of policyholders with car age $2$ and driver age $62$ that had predicted value greater than $5$, the degree of freedom for the chi-squared test was $2$. The p-values from this test are approximately $0.29$ for the DPMM model and $0.32$ for the PYMM model, which are greater than the significance level $0.05$. Therefore, we fail to reject the null hypothesis that the observed distribution in the testing data does not differ significantly from the predictive distribution of the number of claims obtained from the testing data. \par

\begin{figure}[ht!]
    \centering
    \subfloat[DPMM]{\includegraphics[width=.45\textwidth]{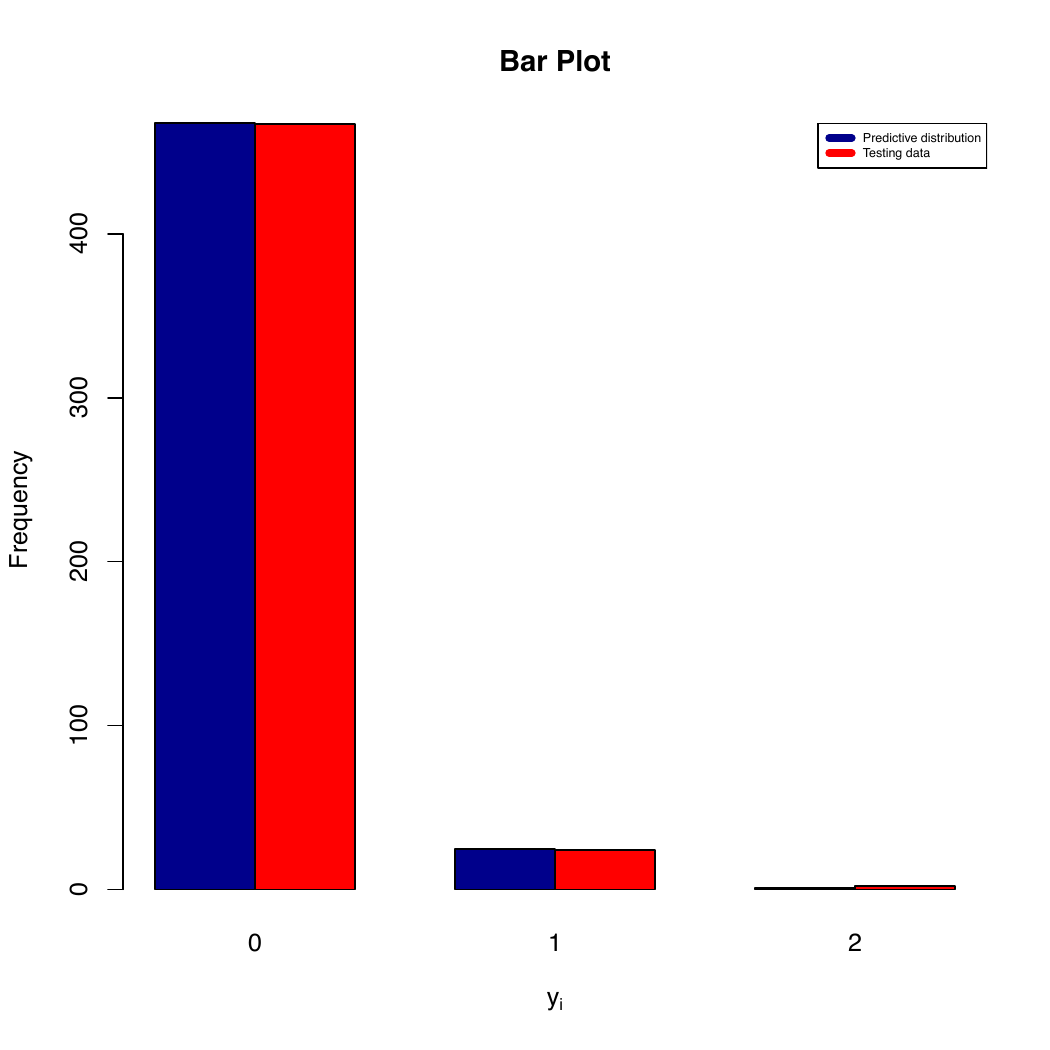}}\quad
    \subfloat[PYMM]{\includegraphics[width=.45\textwidth]{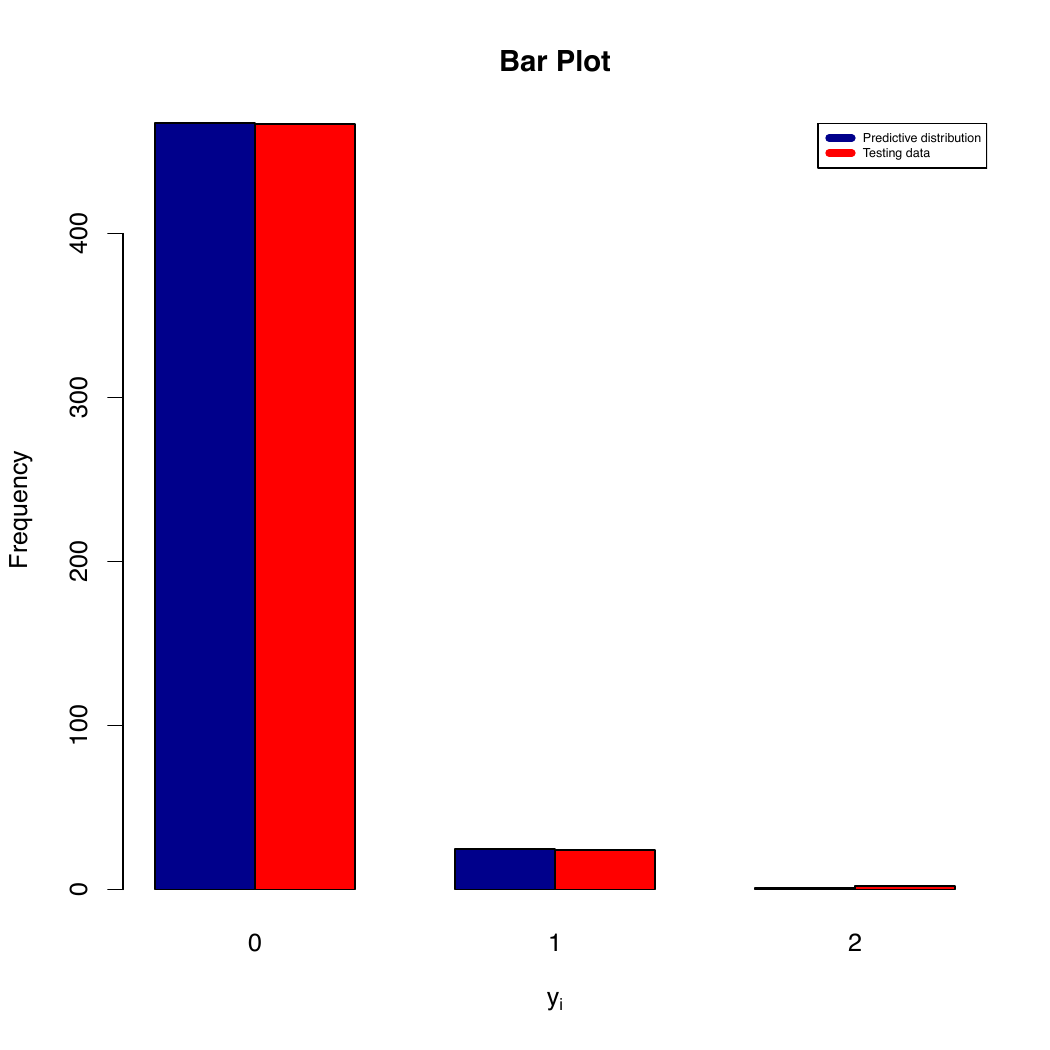}}\quad
    \subfloat[Parametric]{\includegraphics[width=.45\textwidth]{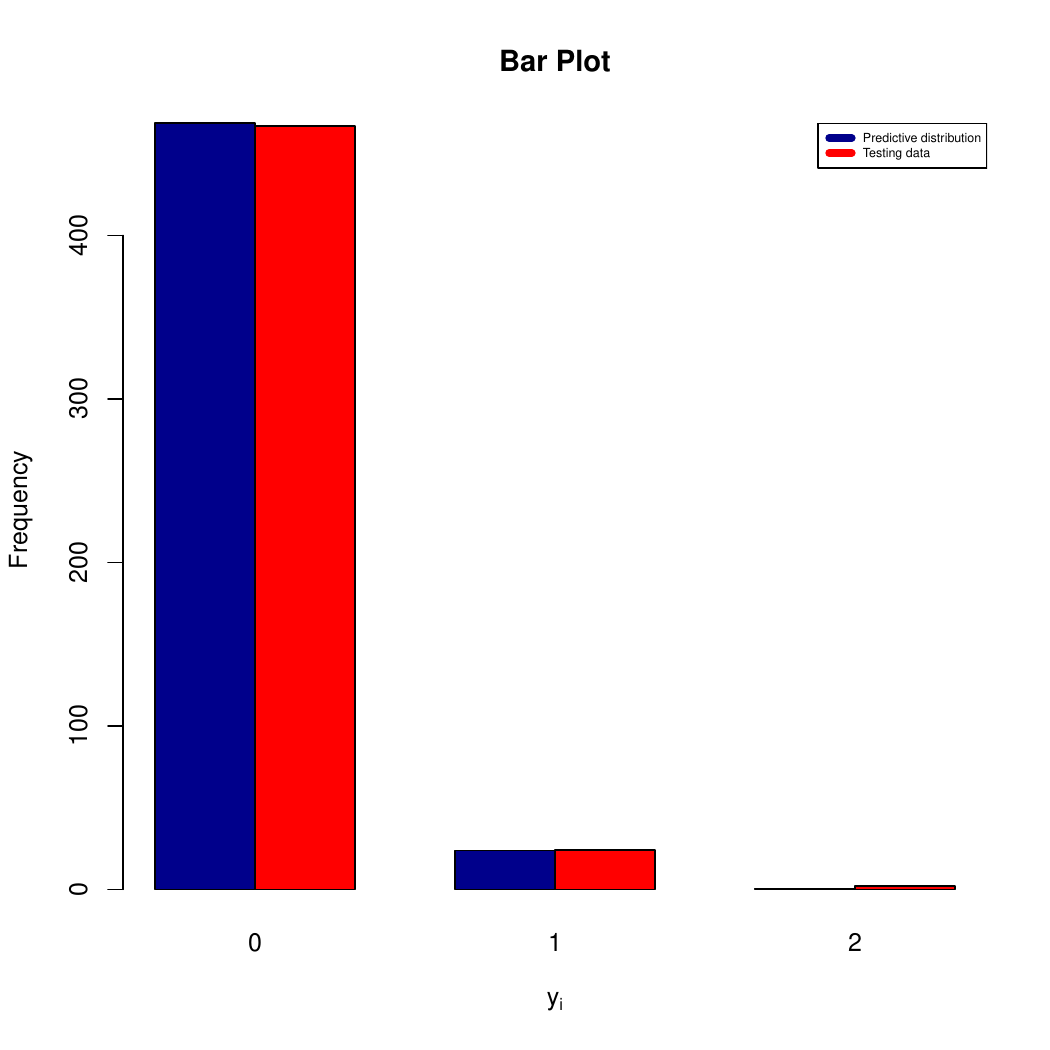}}
    \caption{Predictive distribution versus histogram of testing data 
    for a class of policyholders with car age $2$ and driver age $62$ for the French motor claims frequency dataset.}
    \label{fig:pred-num}
\end{figure}

\indent We also compared our BNP regression models to the classical non-Bayesian parametric Poisson regression. The posterior predictive distribution for a class of policyholders with car age $2$ and driver age $62$ versus the histogram of testing data for this class of policyholders for the DPMM, PYMM, and parametric model are shown in Figure \ref{fig:pred-num}. The plots suggest that our DPMM and PYMM models are able to capture the shape of the testing data set very well.  \par 

\indent In order to assess the prediction performance of the DPMM, PYMM, and parametric model on the testing data, we calculated the mean squared error of the predictions. This is the average squared difference between the predicted number of claims and the observed number of claims in the testing data. The MSE values are shown in Table \ref{tab:mse-num} after rounding to three decimal places. We see that the DPMM and PYMM have similar prediction performance with a slight edge for PYMM, while the parametric model has almost twice the MSE. \par 

\begin{table}[ht!]
    \centering
    \begin{tabular}{|c|c|c|}
    \hline
     DPMM & PYMM & Parametric  \\
     \hline
     $0.828$ & $0.786$ & $1.722$ \\ [1ex]
     \hline
    \end{tabular}
    \caption{MSE of the predictions for the French motor claims frequency dataset.}
    \label{tab:mse-num}
\end{table}

\begin{figure}[ht!]
    \centering
    \subfloat[DPMM]{\includegraphics[width=.45\textwidth]{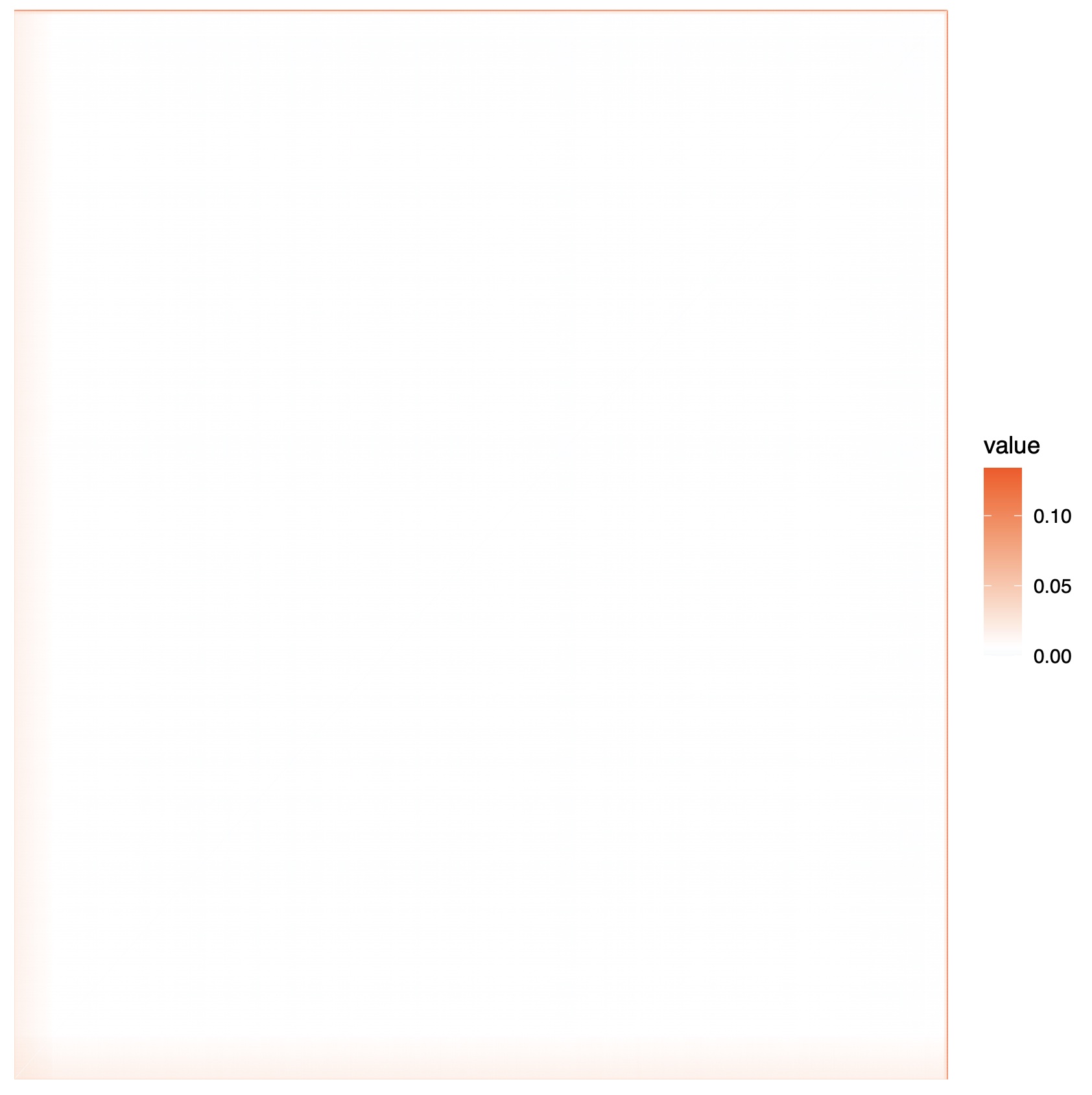}}\quad
    \subfloat[PYMM]{\includegraphics[width=.45\textwidth]{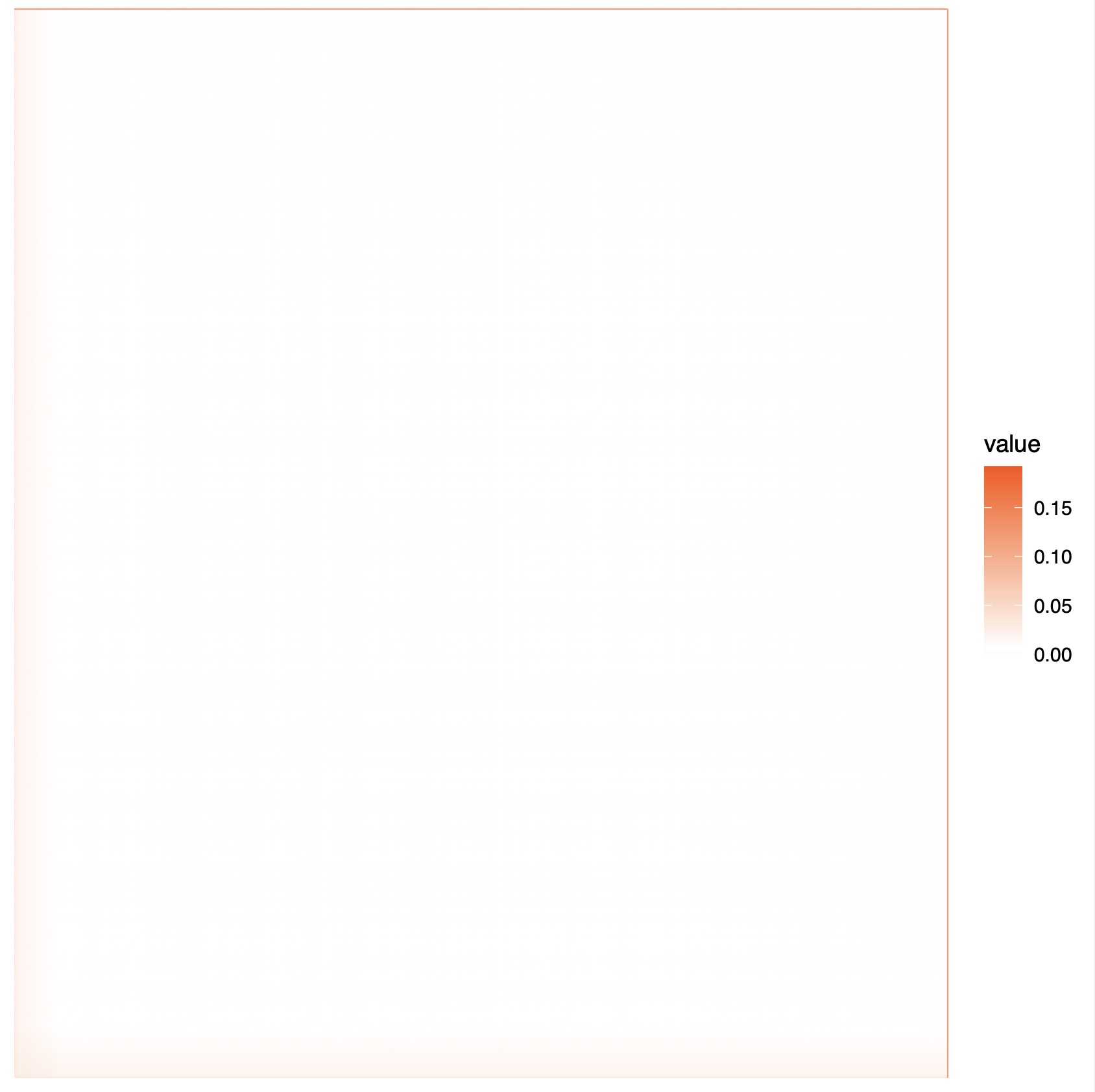}}
    \caption{Heatmap of the dissimilarity matrix showing clustering performance for the French motor claims frequency dataset.}
    \label{fig:heat-num}
\end{figure}

\indent We also checked the posterior clustering performance of our BNP regression models for the French motor insurance claims frequency data. The clustering was done based on samples from the posterior distribution of $\bbet_1,\dots,\bbet_n$. We computed the $n \times n$ dissimilarity matrix based on samples from posterior distribution of $\bbet_1,\dots,\bbet_n$ using the approach described in the Section \ref{sec:clustering}. Heat maps of the dissimilarity matrix for this data are shown in Figure \ref{fig:heat-num}. In these heat maps, observations are on both horizontal and vertical axes and blocks with light colors indicate the clusters. We can see there is approximately $1$ cluster, suggesting that the French motor insurance claims frequencies can be possibly modeled using a single parametric Poisson regression. \par

\section{Modeling Claims Severity}
\indent The second step in developing any insurance pricing model is predicting the claims severity, i.e. the amount of each claim. Let us assume that there are $n$ independent policyholders each with a set of $k$ covariates. The $i$th policyholder's claim amount is denoted by $z_i$ and the risk factor or covariate vector is denoted by $\bm{x}_i = (1,x_{i1},\dots,x_{ik})^T$. The classical non-Bayesian parametric normal regression (multiple linear regression) model that has been widely used to model the log(claims amount) of the $i$th individual is
\begin{align*}
     & \log(z_i) \mid (\bbet,\sigma^2) \ \stackrel{ind.}{\sim} \ \text{Normal}(\bx'_i \bbet , \sigma^2), \quad i=1,\dots,n. \nonumber
\end{align*}
However, this model assumes that the response to covariates is similar for each individual because each data point has the same regression parameter vector $(\bbet,\sigma^2)$, where $\bbet$ is the vector of regression coefficients and $\sigma^2$ is the error variance. Alternatively, the DPMM and PYMM model allow each data point to have its own regression parameter vectors, $(\bbet_i,\sigma_{i}^2)$, for $i = 1,\dots,n$, and because of clustering property of DP prior and PY prior, the data points are clustered by their shared regression parameters.

\subsection{Model Description} \label{sec:bnp-sev}
\indent The BNP regression model parameters here are $\bth_i = (\bbet_i,\sigma_{i}^2)$, for $i = 1,\dots,n$, where $\bbet_i = (\beta_{i0},\beta_{i1},\dots,\beta_{ik})$ is the vector of regression coefficients and $\sigma_{i}^2$ is the error variance associated with the $i$th individual. Note that we consider only policyholders with positive amount of claims and model their claims amount in the log scale. We propose two BNP models for the log(claims amount) data as follows.

First, we set a DP prior on the mixing distribution over the regression parameters, $\bth_i = (\bbet_i,\sigma_{i}^2)$, and we model the amount of claims in log scale using a Dirichlet process mixture of normal regression. Our DPMM model can be written as
\begin{align} \label{eq:dpmm-sev}
     & \log(z_i) \mid (\bbet_i,\sigma_{i}^2) \ \stackrel{ind.}{\sim} \ \text{Normal}(\bx'_i \bm{\beta}_i , \sigma_{i}^2) \nonumber \\
     & \bm{\theta}_i = (\bbet_i,\sigma_{i}^2) \mid G \ \stackrel{iid}{\sim} \ G  \nonumber \\
     & G \ \sim \ \text{DP}(\alpha,G_0) \\
     & G_0 \ = \  G_{0_{\bbet \mid \sigma^2}} \times G_{0_{\sigma^2}} \nonumber \\
     & G_{0_{\bbet \mid \sigma^2}} \ = \ \text{Multivariate-Normal}(\bm{0}_{k+1} , n_{0} \sigma^2 I_{k+1}) \nonumber \\
     & G_{0_{\sigma^2}} \ = \ \text{Inverse-Gamma}(a,b) \nonumber \\
     & \alpha \ \sim \ \text{Gamma}(1,1). \nonumber
\end{align}
The coefficient $n_0$ plays an important role in the covariance matrix of the multivariate normal distribution which needs to be specified carefully by trial and error. After trying out several values for $n_0$ in practice, we found out that $n_0 = 0.5$ works well in this case. $G_{0_{\sigma^2}}$ is the base distribution for $\sigma_{i}^2$ which is chosen to be an inverse gamma distribution with shape parameter $a=3$ and scale parameter $b=5$. 

Alternatively, we can set a PY prior instead of a DP prior on the mixing distribution over the regression parameters, $\bth_i = (\bbet_i,\sigma_{i}^2)$, and model the amount of claims in log scale using a Pitman-Yor process mixture of normal regression. Our PYMM model can be written as
\begin{align} \label{eq:pym-sev}
     & \log(z_i) \mid (\bm{\beta}_i,\sigma_{i}^2) \ \stackrel{ind.}{\sim} \ \text{Normal}(\bx'_i \bm{\beta}_i , \sigma_{i}^2) \nonumber \\
     & \bm{\theta}_i = (\bm{\beta}_i,\sigma_{i}^2) \mid G \ \stackrel{iid}{\sim} \ G  \nonumber \\
     & G \ \sim \ \text{PY}(d,\alpha,G_0) \\
     & G_0 \ = \  G_{0_{\bbet \mid \sigma^2}} \times G_{0_{\sigma^2}} \nonumber \\
     & G_{0_{\bbet \mid \sigma^2}} \ = \ \text{Multivariate-Normal}(\bm{0}_{k+1} , n_{0} \sigma^2 I_{k+1}) \nonumber \\
     & G_{0_{\sigma^2}} \ = \ \text{Inverse-Gamma}(a,b) \nonumber \\
     & d \sim \text{Uniform}(0,1) \nonumber \\
     & (\alpha + d) \mid d \ \sim \ \text{Log-Normal}(0,1). \nonumber
\end{align}

\subsection{Computation of Posterior Predictive Distributions} \label{sec:pred_sev_dpmm_pym}
\indent Henceforth, we use $y_i = \log(z_i)$ to denote the log(claims amount) for the $i$th individual. We are interested in the predictive distribution for the future log(claim amount) $y_{n+1}$. Following similar steps as in the computation of posterior predictive distribution of the DPMM model for claims frequency in the previous sections, we have \\
\begin{dmath*}[style={\small},spread={5pt}]
    f{(y_{n+1} \mid y_1,\dots,y_n)} \approx \frac{1}{T} \sum_{t=1}^{T} \lb \frac{\alpha^{(t)}}{\alpha^{(t)}+n} \rb \int f{(y_{n+1} \mid \bth_{n+1})}G_0(d\bth_{n+1}) 
    + \frac{1}{T} \sum_{t=1}^{T} \lb \frac{1}{\alpha^{(t)}+n} \sum_{j=1}^{n} f{(y_{n+1} \mid \bth_{j}^{(t)})} \rb. \\
\end{dmath*}
Since the base measure $G_0$ in model (\ref{eq:dpmm-sev}) is a conjugate prior for the likelihood given by this model, the integral $\int f(y_{n+1} \mid \bth_{n+1})G_0(d\bth_{n+1})$ above can be calculated in closed form. Therefore, the posterior predictive distribution of the log(claims severity) for the DPMM equals: 
\begin{dmath}[style={\small},spread={5pt}]\label{eq:dpmm-pred-am}
    f{(y_{n+1} \mid y_1,\dots,y_n)} \approx \frac{1}{T} \sum_{t=1}^{T} \lb \frac{\alpha^{(t)}}{\alpha^{(t)}+n} \rb 
    \times \lb \frac{1}{\sqrt{2\pi}} \frac{1}{n_0^{\frac{k+1}{2}}} \frac{b^a}{\Gamma(a)} \frac{1}{\sqrt{det(M)}} \frac{\Gamma(\frac{1}{2}+a)}{\left[b+\frac{1}{2}y_{n+1}^2-\frac{1}{2}\bm{d}'M^{-1}\bm{d}\right]^{(\frac{1}{2}+a)}} \rb
    + \frac{1}{T} \sum_{t=1}^{T} \lb \frac{1}{\alpha^{(t)}+n} \sum_{j=1}^{n} f{(y_{n+1} \mid \bth_j^{(t)})}\rb,
\end{dmath}
where $M=\bx_{n+1}\bx'_{n+1}+\frac{1}{n_0}I$ and $\bm{d}'=\bm{y}_{n+1}\bx'_{n+1}$. \\

\indent Similarly, the posterior predictive distribution of the log(claims severity) for the PYMM equals: \\
\begin{dmath}[style={\small},spread={5pt}] \label{eq:pym-pred-am}
    f{(y_{n+1} \mid y_1,\dots,y_n)} \approx \lb\frac{1}{T} \sum_{t=1}^{T} \frac{\alpha^{(t)}+d^{(t)} K_n^{(t)}}{\alpha^{(t)} + n}\rb
    \times \lb \frac{1}{\sqrt{2\pi}} \frac{1}{n_0^{\frac{k+1}{2}}} \frac{b^a}{\Gamma(a)} \frac{1}{\sqrt{det(M)}} \frac{\Gamma(\frac{1}{2}+a)}{(b+\frac{1}{2}y_{n+1}^2-\frac{1}{2}\bm{d}'M^{-1}\bm{d})^{(\frac{1}{2}+a)}} \rb
    + \frac{1}{T} \sum_{t=1}^{T} \frac{1}{(\alpha^{(t)}+n)} \lb\sum_{j=1}^{n} f{(y_{n+1} \mid \bth_{j}^{(t)})} - d^{(t)} \sum_{j=1}^{K_{n}^{(t)}} f{(y_{n+1} \mid {\bth_j^*}^{(t)})} \rb,
\end{dmath}
where $M=\bx_{n+1}\bx'_{n+1}+\frac{1}{n_0}I$ and $\bm{d}'=\bm{y}_{n+1}\bx'_{n+1}$. 

\subsection{Posterior Sampling} \label{sec:implem-sev}
Note that Neal’s Algorithm 8 can also be used for models with a conjugate prior in order to avoid computing the integrals $\int G_0(\theta) f(y_i \mid \theta) \,d\theta$ for the conditional probabilities in the MCMC algorithm. In the second step of this algorithm, we draw a new value for the distinct parameter $\bm{\phi}_c$ from the posterior distribution of $(\bm{\phi} \mid \bm{y_c})$ when we assume that the base measure $G_0$ is the prior distribution. Since the probability base measure $G_0$ in models (\ref{eq:dpmm-sev}) and (\ref{eq:pym-sev}) is a conjugate prior for the likelihood of the models, we use this conjugacy to implement the Gibbs sampling update where a new value for $\bm{\phi}_c$ is drawn from its posterior distribution given the data associated with cluster label $c$.

\section{Illustration: French Motor Insurance Claims Severity Dataset}
\indent In this section, we consider the freMTPLsev dataset, which contains only nonzero claims amount for $16,181$ motor third-part liability policies (observed mostly during a one year period). This dataset has $2$ columns as follows:
\begin{itemize}
    \item \textit{PolicyID} The policy ID (used to link with the contract dataset).
    \item \textit{ClaimAmount} The cost of the claim, seen as at a recent date.
\end{itemize} 

First, we linked the freMTPLsev dataset with the freMTPLfreq dataset using PolicyID column in order to have risk features in our dataset. We considered the claim amounts in log scale as response variable. Due to computational limit and time, we used two covariates in our BNP regression models: driver age (DriverAge column) and car age (CarAge column) which are ordinal variables and are treated as continuous variables. We ran this model with two covariates in R for $50000$ MCMC iterations on a HPC cluster, and it took $3.06$ days for the DPMM and $2.83$ days for the PYMM to be completed. We randomly selected $10\%$ of the data points as training data and used it to fit our BNP regression models. The remainder was set as testing data to evaluate the fitted models' prediction performance. To avoid overflow or underflow issues in R, we standardized all the covariates. Since we have two covariates here, the risk factor vector is in the form of $\bx_i = (1,x_{i1},x_{i2})^T$, for $i = 1,\dots,n$, and the BNP regression model parameters are $\bth_i = (\bbet_i,\sigma_{i}^2)$, for $i = 1,\dots,n$, where $\bbet_i = (\beta_{i0},\beta_{i1},\beta_{i2})$ is the vector of regression coefficients and $\sigma_{i}^2$ is the error variance for the $i$th individual. Therefore, our Dirichlet process mixture of normal regression model here is:
\begin{align*} 
     & y_i \mid (\bm{\beta}_i,\sigma_{i}^2) \ \stackrel{ind.}{\sim} \ \text{Normal}(\beta_{i0}+\beta_{i1} x_{i1}+\beta_{i2} x_{i2} , \sigma_{i}^2) \nonumber \\
     & \bm{\theta}_i = (\bbet_i,\sigma_{i}^2) \mid G \ \stackrel{iid}{\sim} \ G  \nonumber \\
     & G \ \sim \ \text{DP}(\alpha,G_0) \nonumber \\
     & G_0 \ = \  G_{0_{\bbet \mid \sigma^2}} \times G_{0_{\sigma^2}} \\
     & G_{0_{\bbet \mid \sigma^2}} \ = \ \text{Multivariate-Normal}(\bm{0}_3 , 0.5 \sigma^2 I_3) \\
     & G_{0_{\sigma^2}} \ = \ \text{Inverse-Gamma}(3,5) \\
     & \alpha \ \sim \ \text{Gamma}(1,1).
\end{align*}
Similarly, our Pitman-Yor process mixture of normal regression model here is: 
\begin{align*} 
     & y_i \mid (\bm{\beta}_i,\sigma_{i}^2) \ \stackrel{ind.}{\sim} \ \text{Normal}(\beta_{i0}+\beta_{i1} x_{i1}+\beta_{i2} x_{i2} , \sigma_{i}^2) \nonumber \\
     & \bm{\theta}_i = (\bm{\beta}_i,\sigma_{i}^2) \mid G \ \stackrel{iid}{\sim} \ G  \nonumber \\
     & G \ \sim \ \text{PY}(d,\alpha,G_0) \nonumber \\
     & G_0 \ = \  G_{0_{\bbet \mid \sigma^2}} \times G_{0_{\sigma^2}} \\
     & G_{0_{\bbet \mid \sigma^2}} \ = \ \text{Multivariate-Normal}(\bm{0}_3 , 0.5 \sigma^2 I_3) \\
     & G_{0_{\sigma^2}} \ = \ \text{Inverse-Gamma}(3,5) \\
     & d \sim \text{Uniform}(0,1) \\
     & (\alpha + d) \mid d \ \sim \ \text{Log-Normal}(0,1).
\end{align*} 

\indent We used trace plots and autocorrelation function plots to check the convergence of the Markov chain to stationary distribution. The plots indicated a good convergence. In addition, diagnostic tests of convergence to the equilibrium distribution of the Markov chain provided by the R package ``coda'' did not indicate any issues with convergence.

\begin{figure}[ht!]
    \centering
    \includegraphics[width=.45\textwidth]{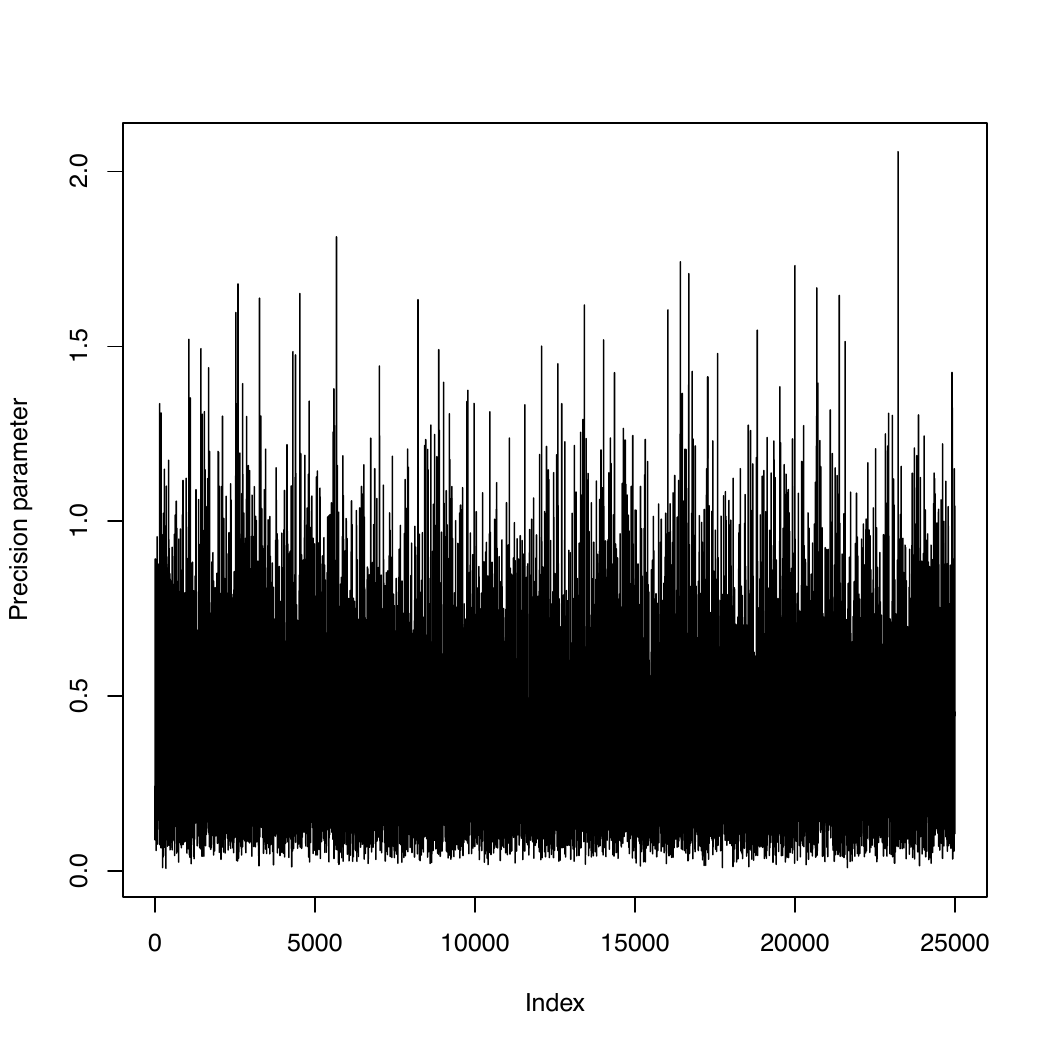}
    \caption{Trace plot of $\alpha$ in DPMM model for the French motor claims severity dataset.}
    \label{fig:alpha-dpmm-am}
\end{figure}

\begin{figure}[ht!]
    \centering
    \subfloat[Trace plot of $\alpha$]{\includegraphics[width=.45\textwidth]{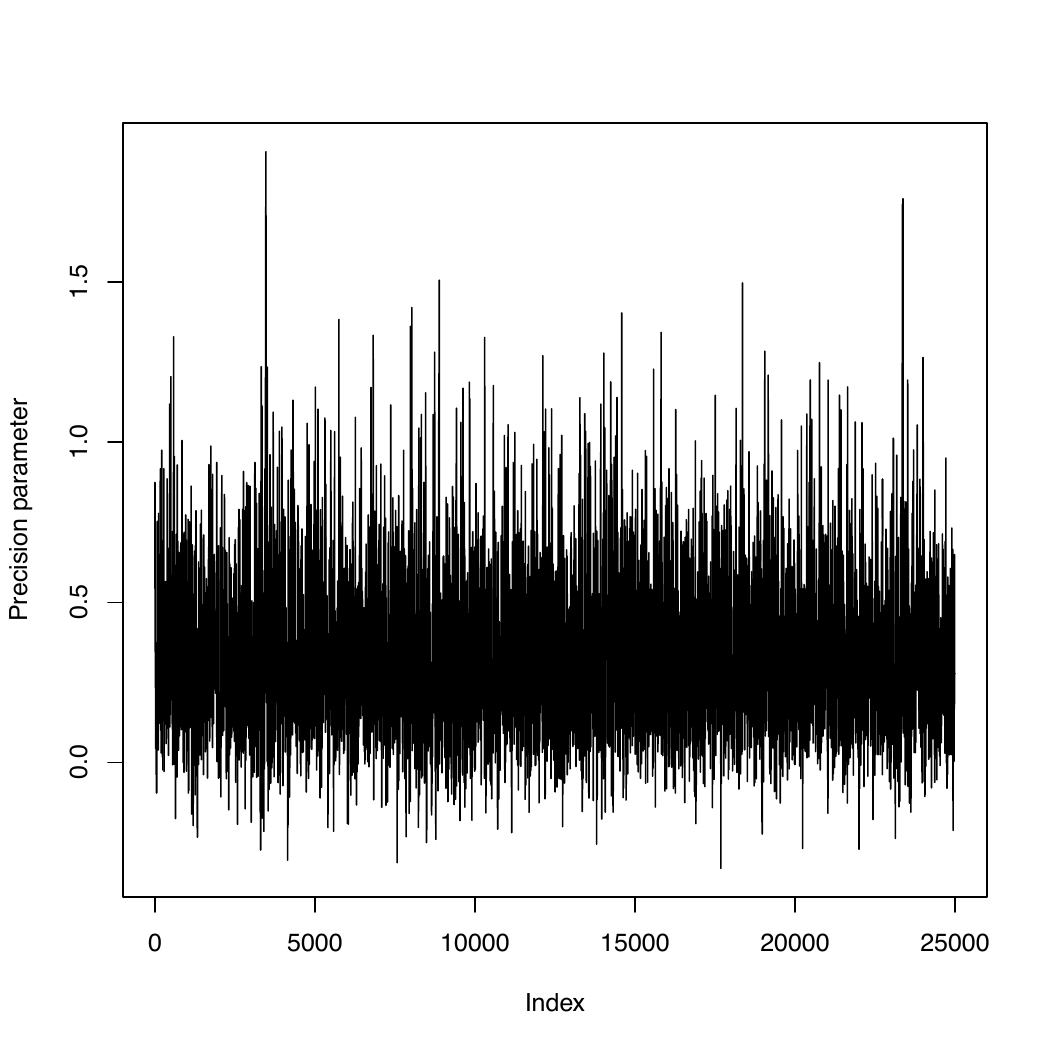}}\quad
    \subfloat[Trace plot of $d$]{\includegraphics[width=.45\textwidth]{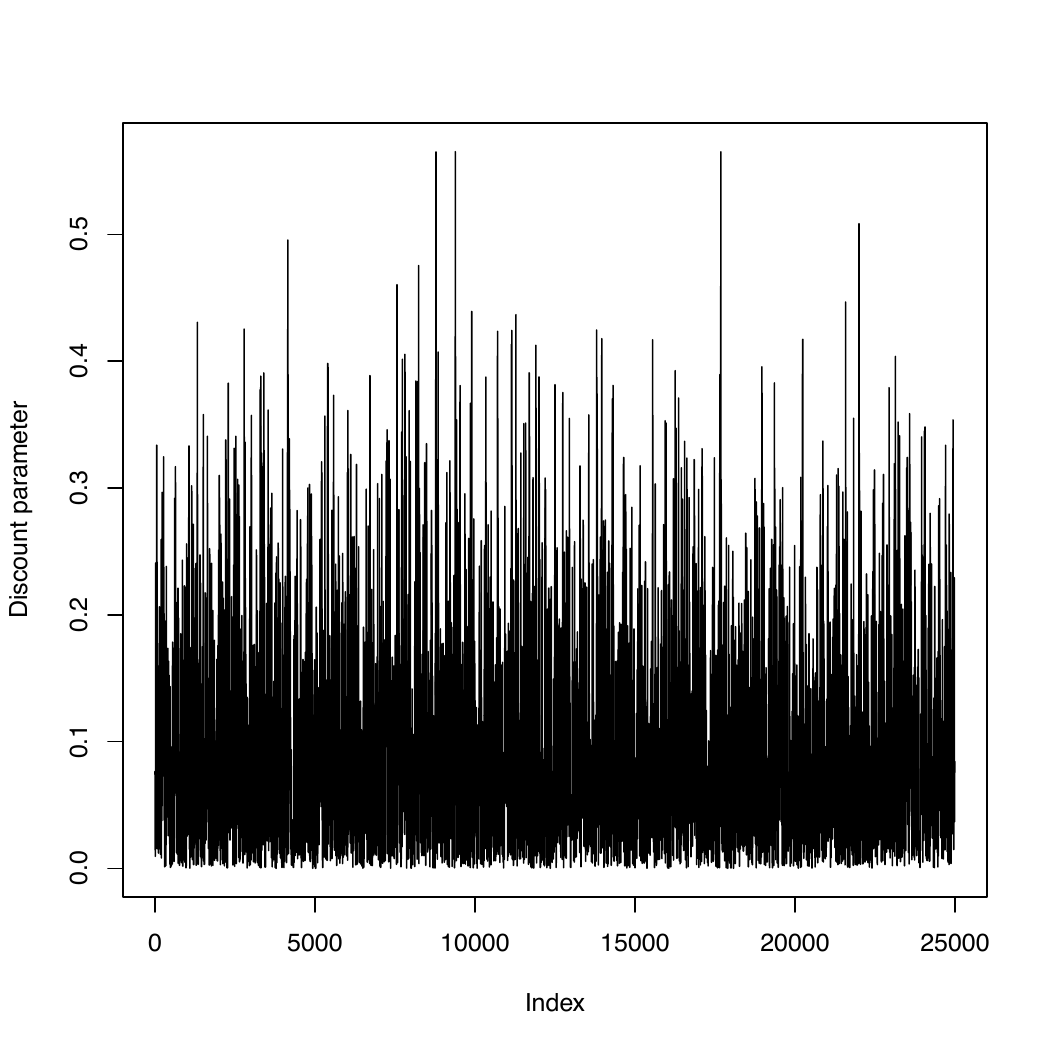}}
    \caption{Trace plots of $\alpha$ and $d$ in PYMM model for the French motor claims severity dataset.}
    \label{fig:alphad-pym-am}
\end{figure}

\begin{figure}[ht!]
    \centering
    \subfloat[DPMM]{\includegraphics[width=.45\textwidth]{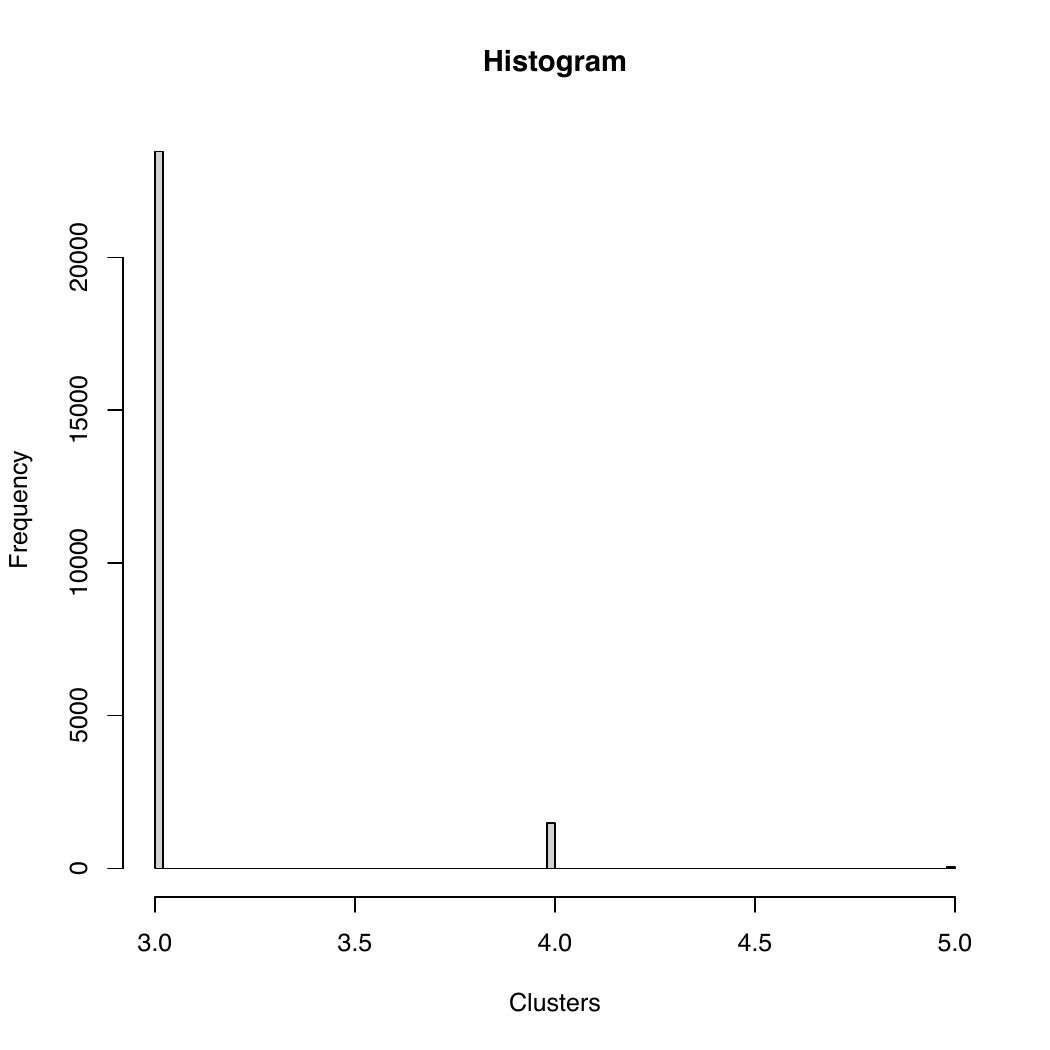}}\quad
    \subfloat[PYMM]{\includegraphics[width=.45\textwidth]{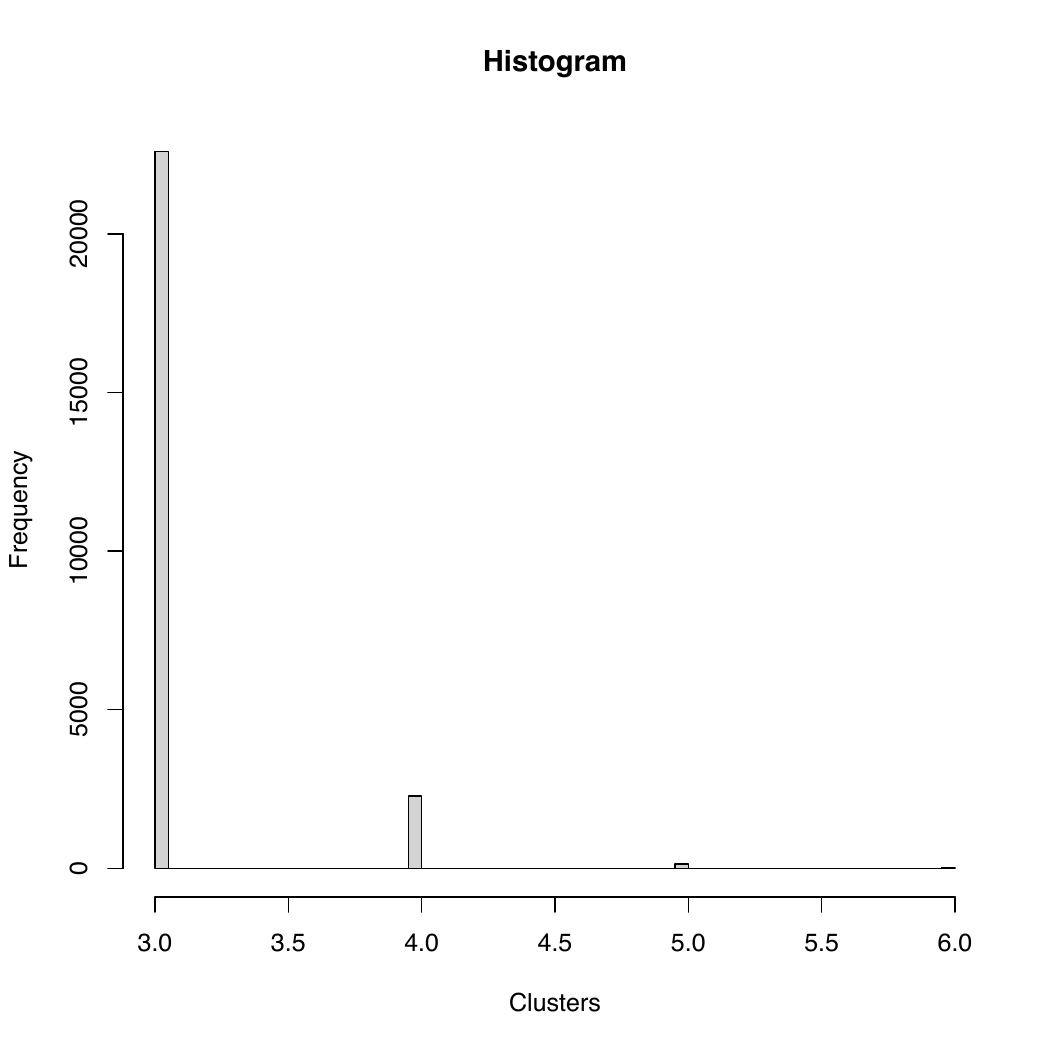}}
    \caption{Histogram of the number of distinct components for the French motor claims severity dataset.}
    \label{fig:hist-am}
\end{figure}

We ran the algorithm for $50000$ iterations, with a burn-in of first $25000$ iterations. The trace plot of $\alpha$ in the DPMM model and trace plots of $\alpha$ and $d$ in the PYMM model are shown in Figures \ref{fig:alpha-dpmm-am} and \ref{fig:alphad-pym-am}. Histograms of the number of distinct components are shown in Figure \ref{fig:hist-am}. \par 

\begin{figure}[ht!]
    \centering
    \subfloat[DPMM]{\includegraphics[width=.45\textwidth]{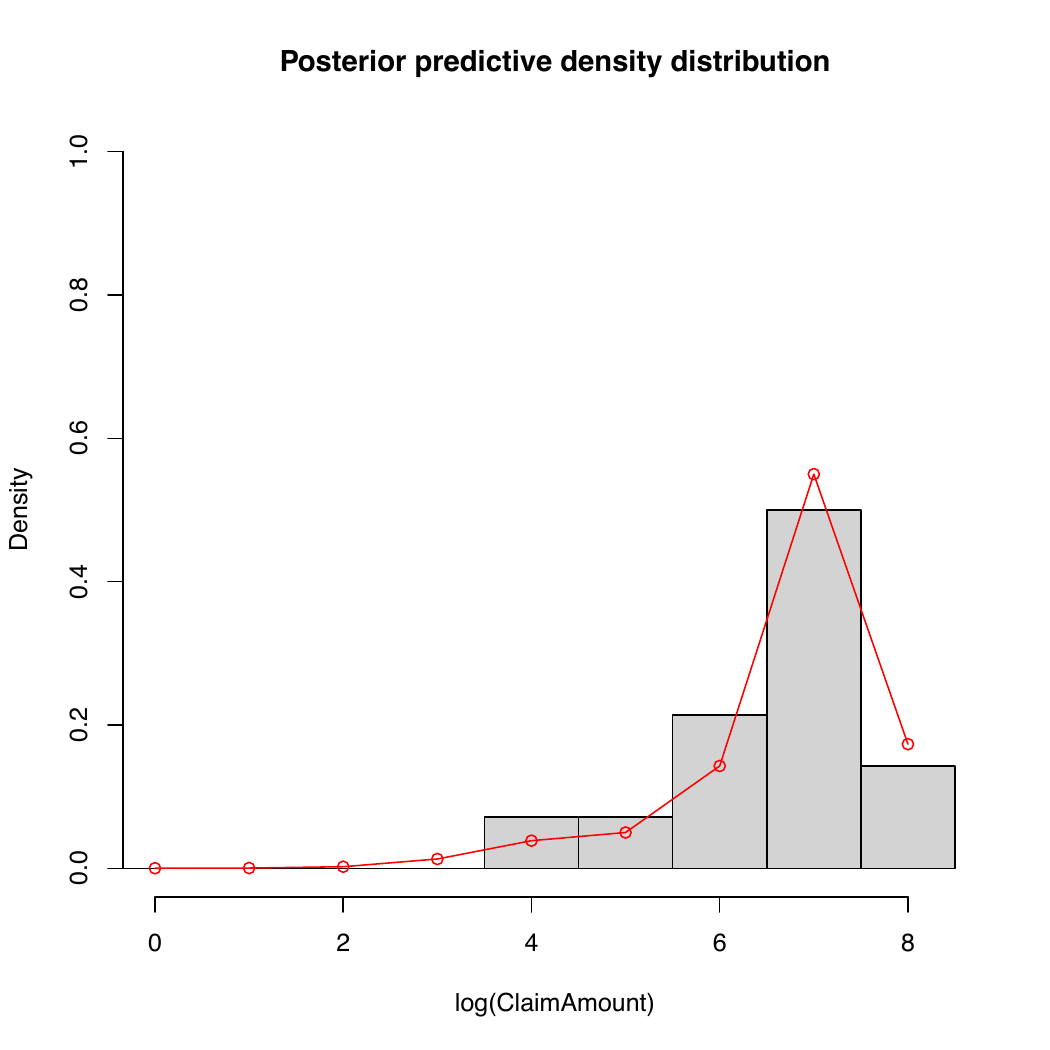}}\quad
    \subfloat[PYMM]{\includegraphics[width=.45\textwidth]{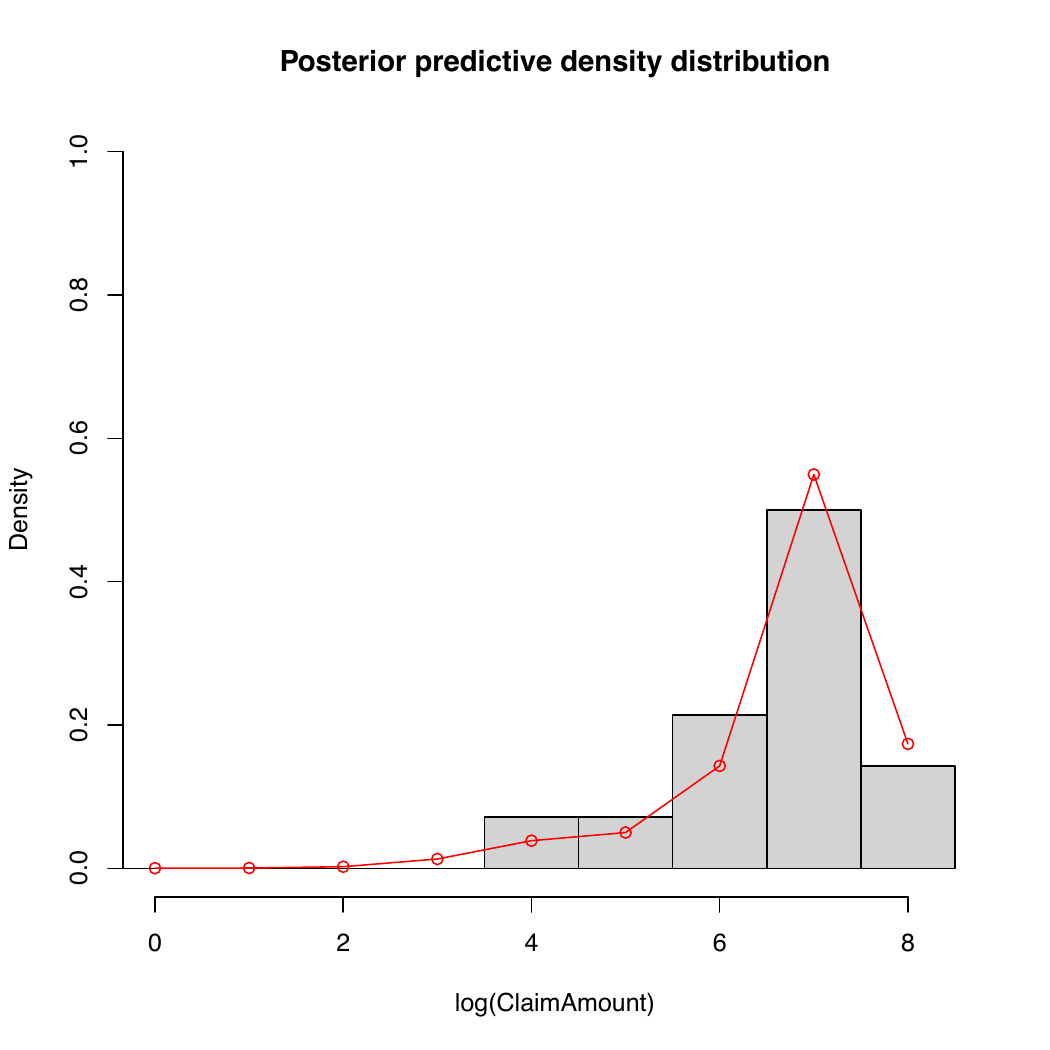}}\quad
    \subfloat[Parametric]{\includegraphics[width=.45\textwidth]{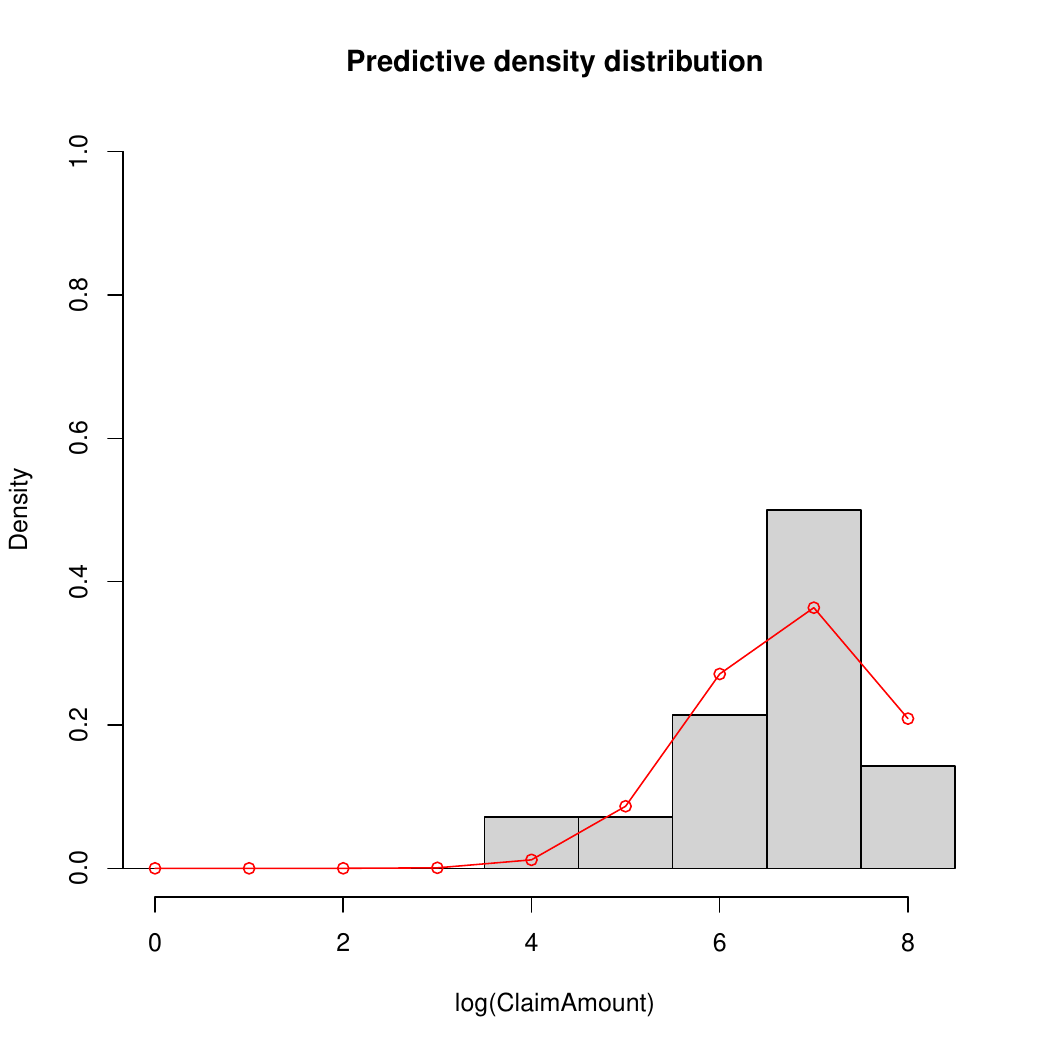}}
    \caption{Posterior predictive density estimate versus histogram of the testing data
    for a class of policyholders with the car age $9$ and the driver age $38$ for the French motor claims severity dataset.}
    \label{fig:pred-am}
\end{figure}

\indent We calculated the predictive density of the log(claims amount) for our BNP regression models using Equations (\ref{eq:dpmm-pred-am}) and (\ref{eq:pym-pred-am}). Also, we compared our BNP regression models to the classical non-Bayesian parametric normal regression (multiple linear regression model). The posterior predictive density estimate plot for a class of policyholders with the car age group $9$ and the driver age group $38$ using the DPMM, PYMM, and parametric model versus the histogram of testing data for this class of policyholders are shown in Figure \ref{fig:pred-am}. The plots suggest that our BNP regression models are able to capture the shape of the testing data set quite well, while the parametric model fails substantially. 

The MSE of the predictions obtained using the three models are shown in Table \ref{tab:mse-am}. Again, we see that DPMM and PYMM have similar prediction performance in this real data application with a slight advantage for PYMM, while the parametric model performs substantially worse. 
\begin{table}[ht!]
    \centering
    \begin{tabular}{|c|c|c|}
    \hline
     DPMM & PYMM & Parametric  \\
     \hline
     $0.223$ & $0.222$ & $0.652$ \\ [1ex]
     \hline
    \end{tabular}
    \caption{MSE of the predictions for the French motor claims severity dataset.}
    \label{tab:mse-am}
\end{table}

\begin{figure}[ht!]
    \centering
    \subfloat[DPMM]{\includegraphics[width=.45\textwidth]{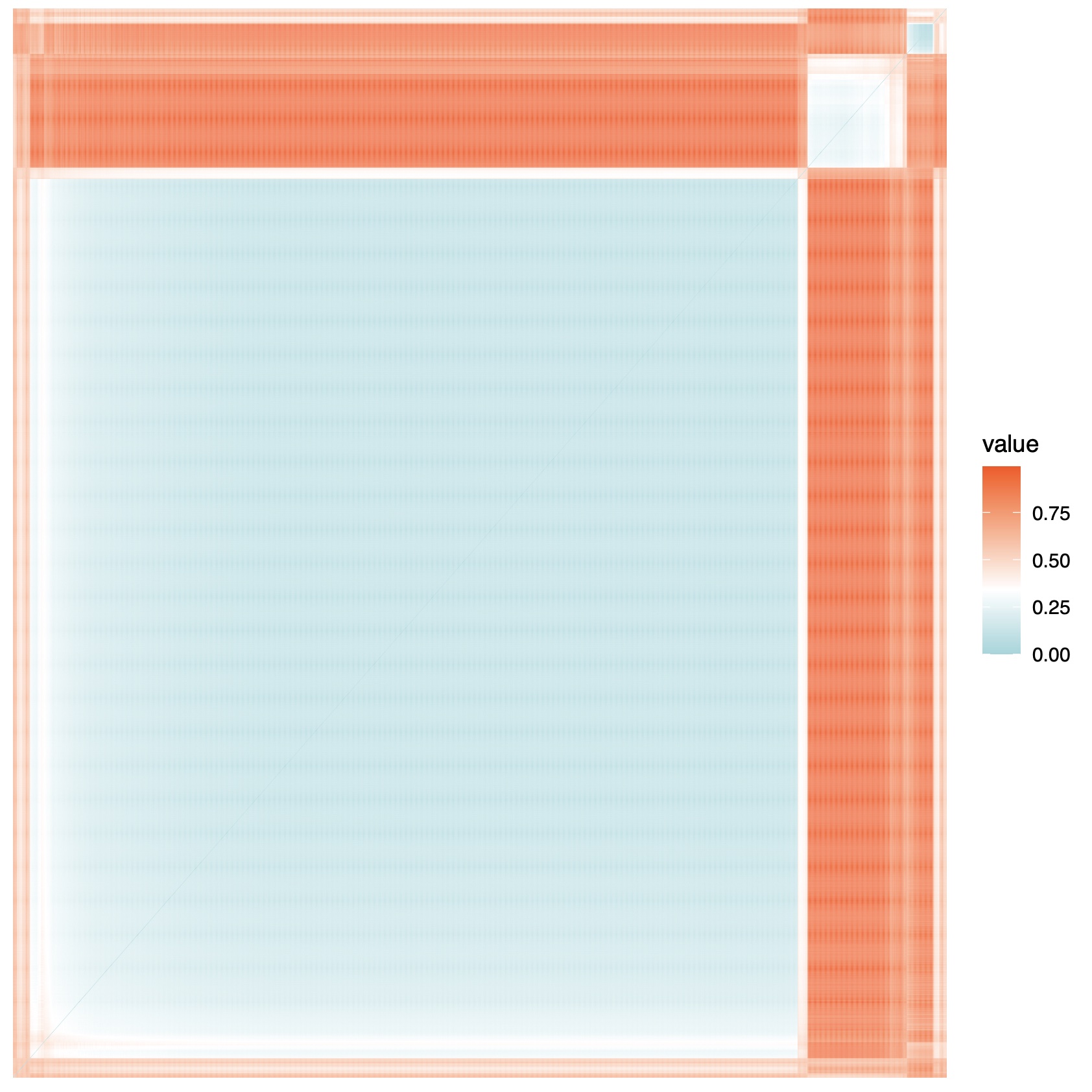}}\quad
    \subfloat[PYMM]{\includegraphics[width=.45\textwidth]{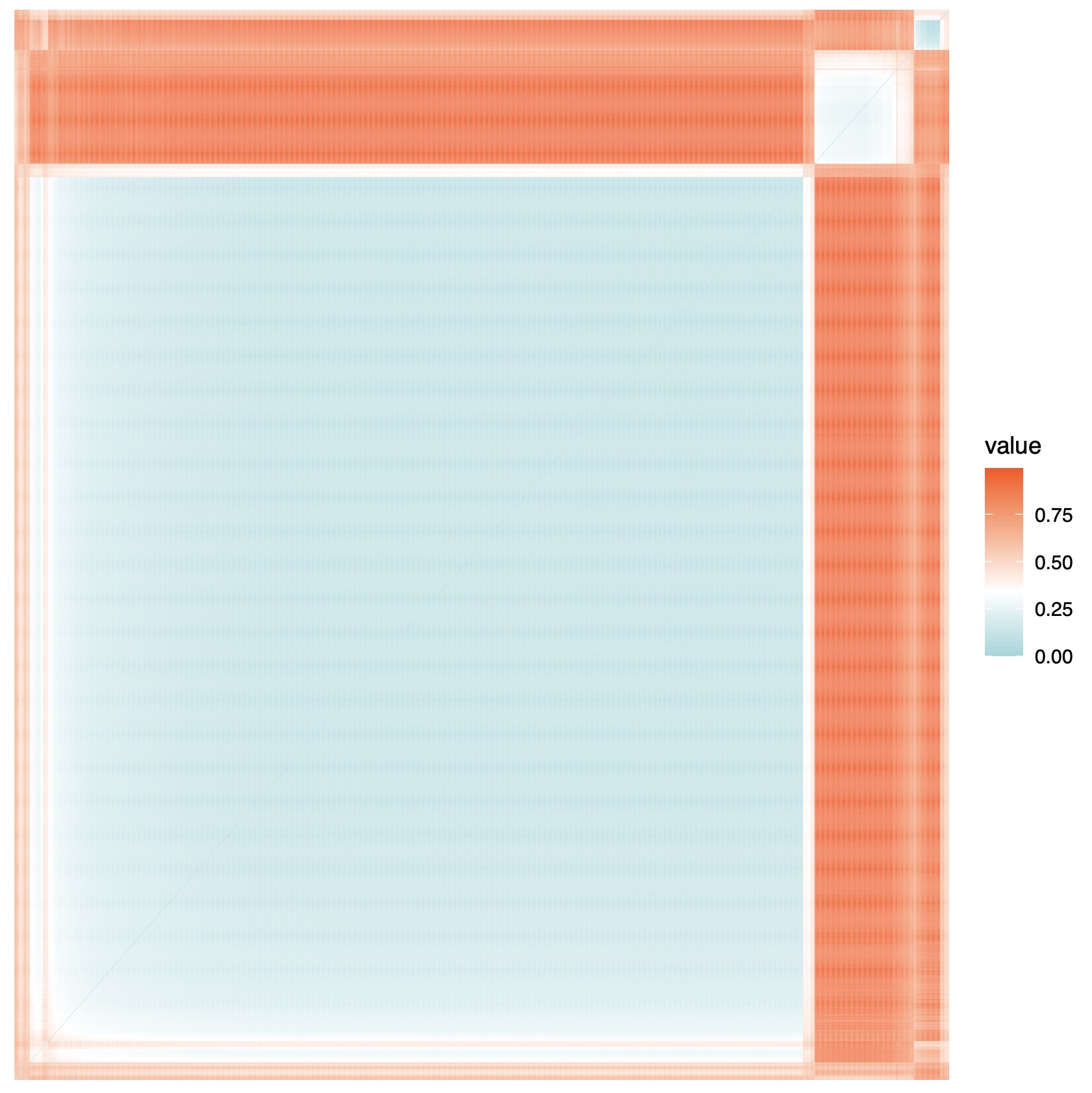}}
    \caption{Heatmap of the dissimilarity matrix showing clustering performance for the French motor claims severity dataset.}
    \label{fig:heat-am}
\end{figure}

\begin{figure}[ht!]
    \centering
    \subfloat[DPMM]{\includegraphics[width=.45\textwidth]{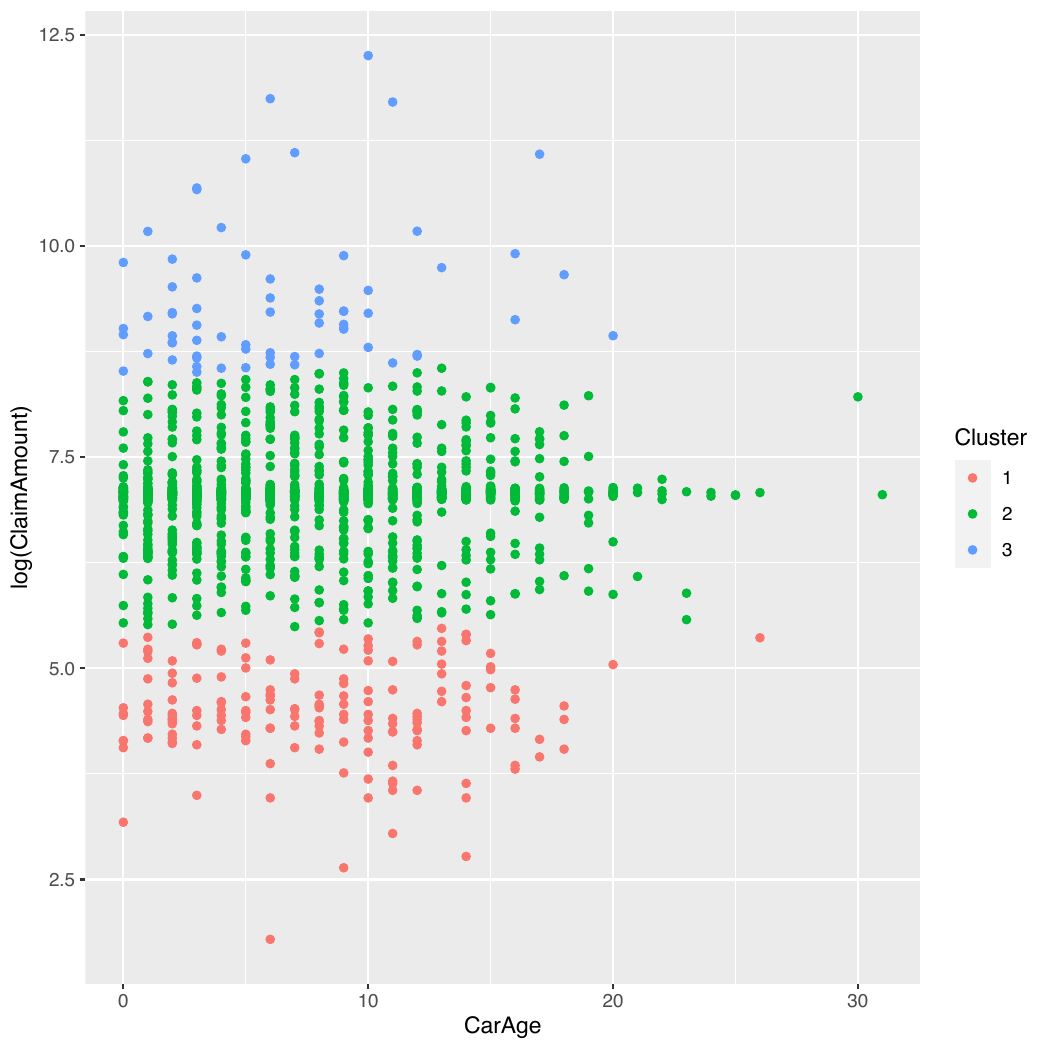}}\quad
    \subfloat[PYMM]{\includegraphics[width=.45\textwidth]{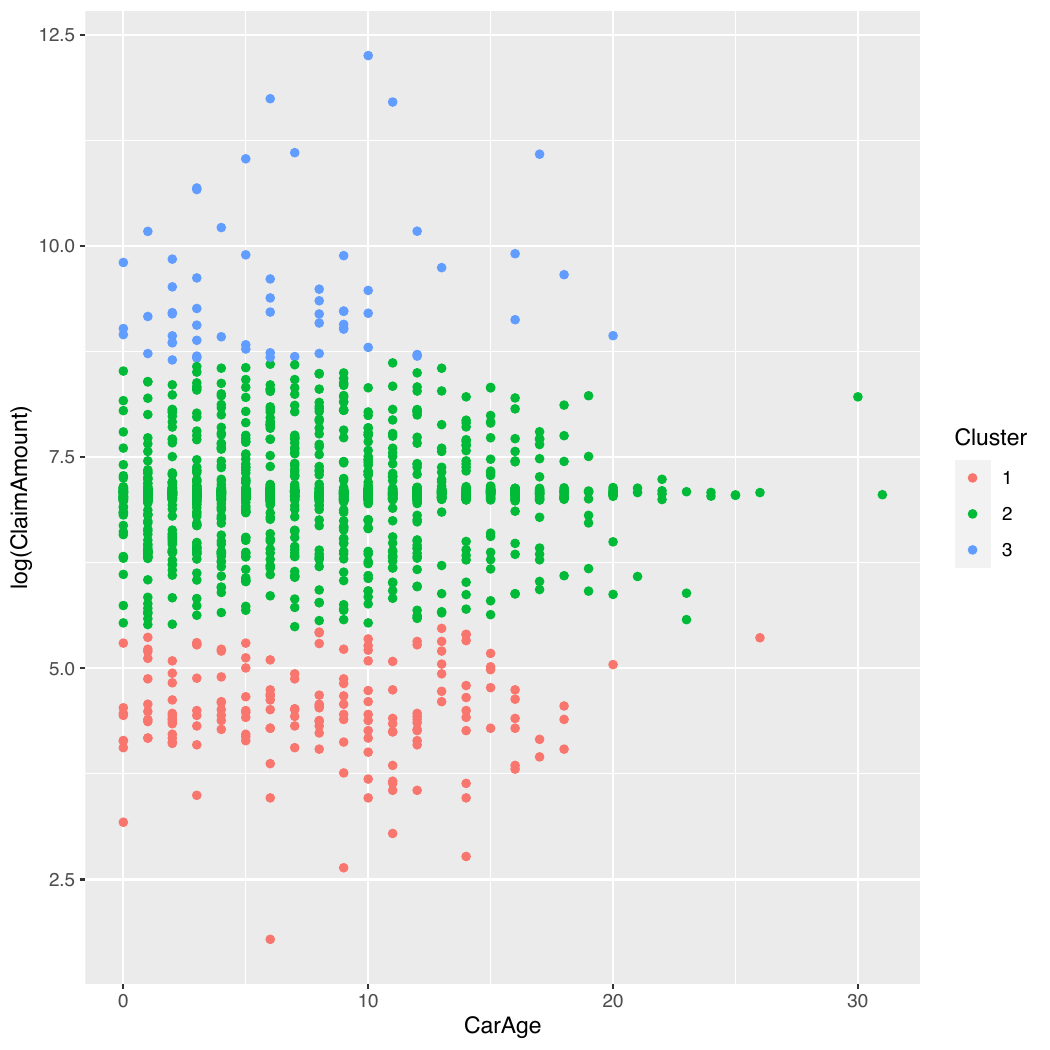}}
    \caption{Scatter plot of log(claims amount) versus car age for the French motor claims severity dataset.}
    \label{fig:car}
\end{figure}

\begin{figure}[ht!]
    \centering
    \subfloat[DPMM]{\includegraphics[width=.45\textwidth]{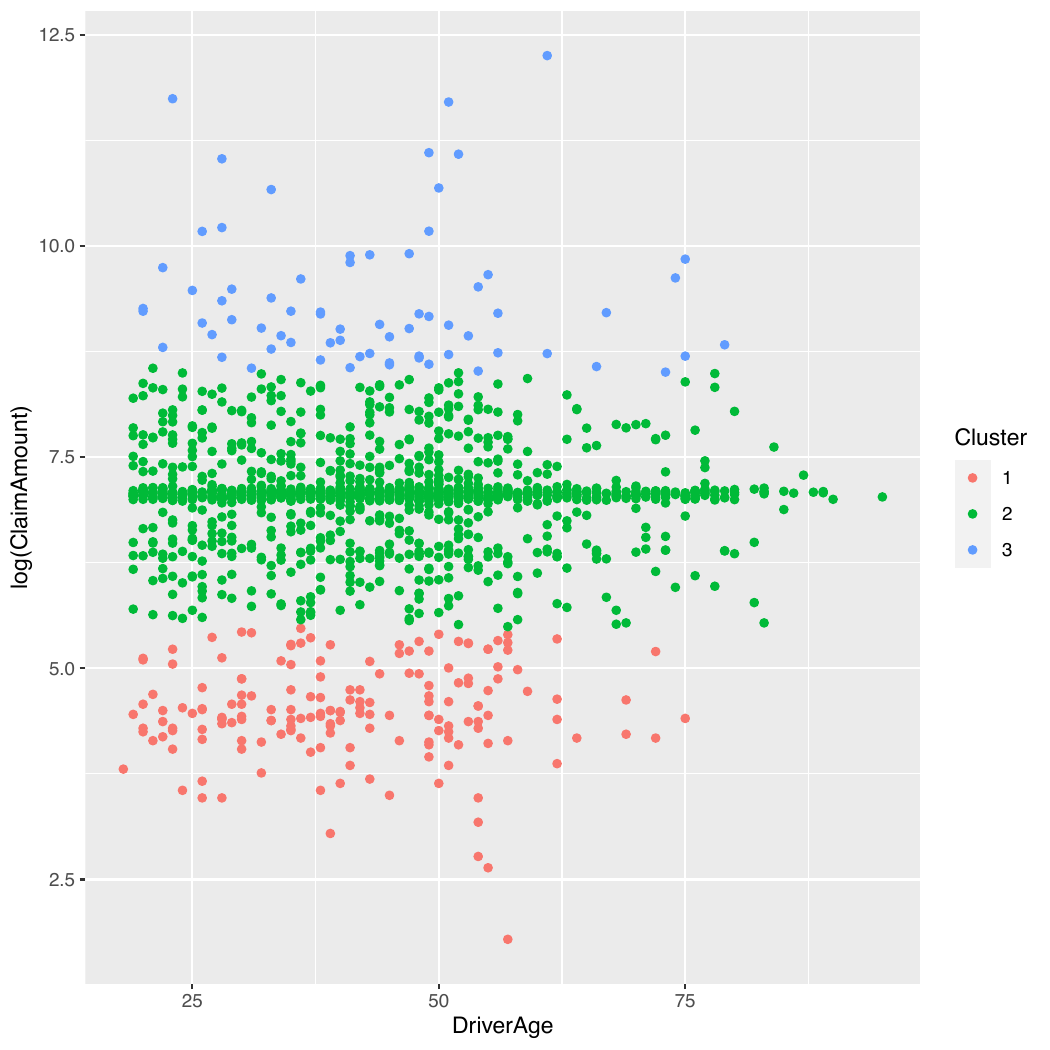}}\quad
    \subfloat[PYMM]{\includegraphics[width=.45\textwidth]{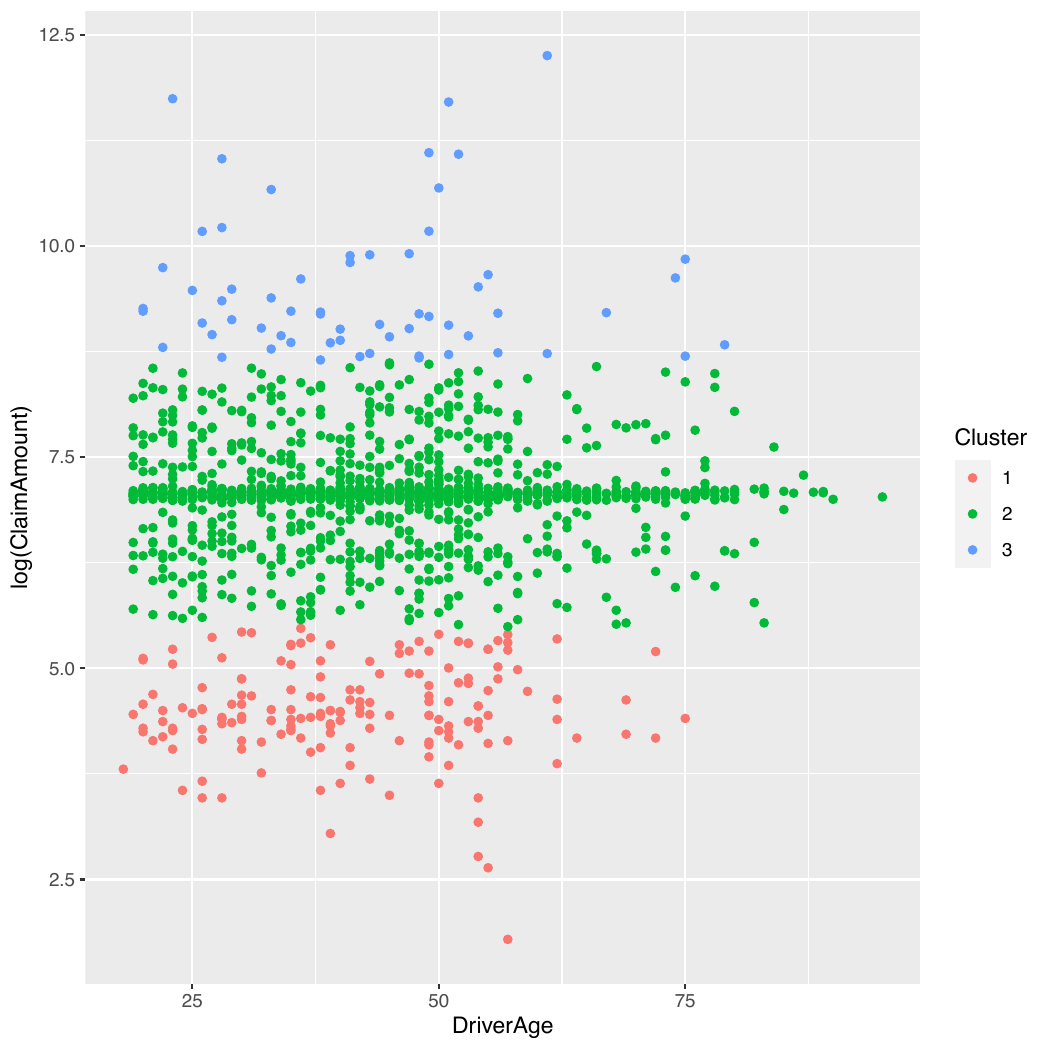}}
    \caption{Scatter plot of log(claims amount) versus driver age for the French motor claims severity dataset.}
    \label{fig:driver}
\end{figure}

\indent Heat maps of the dissimilarity matrix for the French motor insurance claims severity data based on the two BNP models are shown in Figure \ref{fig:heat-am}. In these heat maps, blocks along the diagonal with blue or white colors indicate clusters. The heat maps show that there are approximately $3$ clusters with different areas, indicating that the French motor insurance claims severity data can be possibly modeled using a mixture of three normal regression models with different weights. The clustering structure of our BNP regression models are also shown in the scatter plots of claims amount in log scale versus car age and driver age based on clusters in Figures \ref{fig:car} and \ref{fig:driver}. The first cluster (in red) are those with log(claims amount) approximately less than $5.25$, the second cluster (in green) are those with log(claims amount) approximately greater than $5.25$ and less than $8.50$, and the last cluster (in blue) are those with log(claims amount) approximately greater than $8.50$. 

\section{Conclusions and future directions}
Based on the findings in this article, we conclude that our BNP regression models are able to capture the shape of the testing data set very well, show a very good prediction performance on the testing data, and are able to find the clustering structure of data. By checking the prediction performance via the mean squared error (MSE) of the predictions and the posterior predictive density estimate plots, PYMM and DPMM are seen to have similar prediction performance with a slight edge for PYMM. The reason for this similarity is that the distribution of data here is not heavy-tailed, as seen in the histograms of testing data in Figure \ref{fig:pred-num} for claims frequency and in Figure \ref{fig:pred-am} for claims severity. When data have distribution with heavier tail than exponential, PYMM will possibly perform better than DPMM. We also compared our BNP regression models to the classical non-Bayesian parametric Poisson regression model and normal regression model (multiple linear regression model). By comparing the MSE of the predictions and the predictive density estimate plots, we see that our DPMM and PYMM models have a much better prediction performance than the corresponding parametric regression models. \par

\indent We have coded DPMM for claims frequency, DPMM for log(claims severity), PYMM for claims frequency, and PYMM for log(claims severity) in R, and the codes are available upon request. Each code consists of the model's MCMC sampling algorithm, its posterior predictive distribution computation, and evaluating its clustering performance. We ran each R code on a HPC cluster, and it took $3.19$ days for the DPMM of claims frequency, $2.91$ days for the PYMM of claims frequency, $3.06$ days for the DPMM of log(claims severity), and $2.83$ days for the PYMM of log(claims severity) to be completed for $50000$ MCMC iterations. The Neal (2000)'s Algorithm 8 can't be run in parallel because each iteration depends on the past iterations in this method. However, we have run the posterior prediction distribution calculations in parallel and this made it considerably faster.\par     

\indent As mentioned above, although the Neal (2000)’s Algorithm 8 Gibbs sampling approach works very well in sampling from the posterior distribution of the conjugate and nonconjugate Dirichlet process mixture models and Pitman-Yor process mixture models, it can be slow to converge and mix poorly specially in a nonconjugate BNP model. When the distinct mixture components in a BNP nonconjugate model are nearby or overlapping, these Gibbs sampling methods can be stuck in local modes and mix poorly, which leads to slow convergence to the true posterior distribution. \cite{jain-neal2007} have developed a new split-merge MCMC technique for DPMM and PYMM models which avoids the problem of being trapped in local modes, allowing the posterior distribution to be fully explored. This split-merge MCMC technique, which has been developed for both conjugate and nonconjugate DPMM and PYMM models, escapes such local modes by splitting or merging the mixture components \cite[see][]{jain-neal2007,jain-neal2004}. This technique has also been commented on and discussed in \cite{dahlcomment}, \cite{robertcomment}, \cite{MacEacherncomment}, and \cite{rejoinder}. For the future work, one can check the performance of the \cite{jain-neal2007} split-merge MCMC technique for sampling from the posterior distribution of DPMM and PYMM models in modeling insurance loss data. \par

\bibliographystyle{tfcad}
\bibliography{ref}


\end{document}